\documentclass[preprint,11pt]{aastex}

\newcommand{\kms}{~km~s$^{-1}$} 
\newcommand{\teff}{$T_{\rm eff}$}
\newcommand{\logg}{$\log g$}

\newcommand{\vt}{$v_{\rm micro}$}

\newcommand{\CS}{CS~29526--110} 
\newcommand{\LP}{LP~706--7} 
\newcommand{\obja}{SDSS 0036--10} 
\newcommand{\objb}{SDSS 2047+00} 
\newcommand{\objc}{SDSS 0126+06} 
\newcommand{\objd}{SDSS 0817+26} 
\newcommand{\obje}{SDSS 0924+40} 
\newcommand{\objf}{SDSS 1707+58} 
\shorttitle{Carbon-Enhanced Metal-Poor Stars. III}
\shortauthors{Aoki et al.}

\begin{document}

\title{Carbon-Enhanced Metal-Poor Stars. III. Main-Sequence Turn-Off
  Stars from the SDSS/SEGUE Sample\altaffilmark{1}}

\author{Wako Aoki\altaffilmark{2,3}, Timothy C. Beers\altaffilmark{4},
Thirupathi Sivarani\altaffilmark{4}, Brian Marsteller\altaffilmark{4,5}, Young Sun
Lee\altaffilmark{4}, Satoshi Honda\altaffilmark{2,6}, John E. Norris\altaffilmark{7}, Sean
G. Ryan\altaffilmark{8}, Daniela Carollo\altaffilmark{9,10}}

\altaffiltext {1}{Based on data collected at the Subaru Telescope,
which is operated by the National Astronomical Observatory of Japan.}

\altaffiltext{2}{National Astronomical Observatory, Mitaka, Tokyo,
181-8588 Japan; email: aoki.wako@nao.ac.jp}
\altaffiltext{3}{Department of Astronomical Science, The Graduate
  University of Advansed Stidies, Mitaka, Tokyo, 181-8588 Japan}
\altaffiltext{4}{Department of Physics and Astronomy, CSCE: Center for
  the Study of Cosmic Evolution, and JINA: Joint Institute for Nuclear
  Astrophysics, Michigan State University, East Lansing, MI
  48824-1116; email: beers@pa.msu.edu, marsteller@pa.msu.edu,
thirupathi@pa.msu.edu}
\altaffiltext{5}{present address: Department of Physics \& Astronomy,
University of California, Irvine, Irvine, CA 92697-4575; email:
marsteller@pa.msu.edu}
\altaffiltext{6}{present address: Gunma Astronomical Observatory,
Takayama, Agatsuma, Gunma 377-0702, Japan; honda@astron.pref.gunma.jp}
\altaffiltext{7}{Research School of Astronomy and Astrophysics, The
Australian National University, Mount Stromlo Observatory, Cotter
Road, Weston, ACT 2611, Australia; email: jen@mso.anu.edu.au}
\altaffiltext{8}{Centre for Astrophysics Research,
  STRI and School of Physics, Astronomy and Mathematics, University of
  Hertfordshire, College Lane, Hatfield AL10 9AB, United Kingdom;
  email: s.g.ryan@herts.ac.uk}
\altaffiltext{9}{INAF -- Osservatorio Astronomico di Torino, 10025 Pino Torinese, Italy}
\altaffiltext{10}{present address:
Research School of Astronomy and Astrophysics, The
Australian National University, Mount Stromlo Observatory, Cotter
Road, Weston, ACT 2611, Australia; email: carollo@mso.anu.edu.au}

\begin{abstract} 

The chemical compositions of seven Carbon-Enhanced Metal-Poor (CEMP)
turn-off stars are determined from high-resolution spectroscopy. Five
of them are selected from the SDSS/SEGUE sample of metal-poor stars.
Another star, also chosen from the SDSS/SEGUE sample, has only a weak
upper limit on its carbon abundance obtained from the high-resolution
spectrum. The effective temperatures of these objects are all higher
than 6000~K, while their metallicities, parametrized by [Fe/H], are all below
$-2$; the star with the lowest iron abundance in this study has [Fe/H] = $-3.1.$
Six of our program objects exhibit high abundance ratios of barium ([Ba/H] $>
+1$), suggesting large contributions of the products of former AGB companions
via mass transfer across binary systems. One star in our study ({\objf})
exhibits a rapid variation in its radial velocity, which is a strong signature
that this star belongs to a close binary. Combining our results with previous
studies provides a total of 20 CEMP main-sequence turn-off stars for which the
abundances of carbon and at least some neutron-capture elements are determined.
Inspection of the [C/H] ratios for this sample of CEMP turn-off stars show that
they are generally higher than those of CEMP giants; their dispersion in this
ratio is also smaller. We take these results to indicate that the
carbon-enhanced material provided from the companion AGB star is preserved at
the surface of turn-off stars with no significant dilution, which appears
counter to expectations if processes such as thermohaline mixing have operated
in unevolved CEMP stars. In contrast to the behavior of [C/H], a large
dispersion in the observed [Ba/H] is found for the sample of CEMP turn-off stars,
suggesting that the efficiency of the s-process in very metal-poor AGB stars may
differ greatly from star to star. Four of the six stars from the SDSS/SEGUE
sample exhibit kinematics that are associated with membership in the outer-halo
population, a remarkably high fraction.


\end{abstract}

\section{Introduction}\label{sec:intro}

Abundance studies of very metal-poor (VMP; [Fe/H] $ <
-2.0$)\footnote{[A/B] = $\log(N_{\rm A}/N_{\rm B})- \log(N_{\rm
    A}/N_{\rm B})_{\odot}$, and $\log\epsilon_{\rm A} = \log(N_{\rm
    A}/N_{\rm H})+12$ for elements A and B.} stars have been pursued
over the past few decades in order to constrain models of
nucleosynthesis, stellar evolution, and early chemical enrichment in
the Galaxy \citep[e.g., ][]{beers05}. One important result of these
studies is the discovery of Carbon Enhanced Metal-Poor (CEMP) stars,
which appear with increasing frequency at lower metallicity
\citep{beers92, beers05, lucatello06, marsteller07}. These stars may
be closely related to carbon stars in the Galactic halo, known as CH
stars \citep{keenan42} and subgiant CH stars \citep{bond74}.

Recent chemical abundance studies
for CEMP stars have revealed that most (70--80\%) CEMP stars also
exhibit excesses of s-process elements such as Ba (the CEMP-s stars,
according to Beers \& Christlieb 2005), indicating that the origin of
the carbon excesses in these stars is likely to be the
triple-$\alpha$ reaction in Asymptotic Giant Branch (AGB) stars
\citep[e.g., ][]{aoki07}. The CEMP stars that are observed at present
are likely to have been polluted by the transfer of carbon-enhanced
material from a (former) AGB companion across a binary system, while
the AGB star itself has now evolved to an unseen white dwarf
\citep[e.g.,][]{lucatello05}. Thus, the abundance patterns of heavy
elements in these stars provide useful constraints on models for
s-process nucleosynthesis in AGB stars. On the order of 20\% of CEMP
stars exhibit no significant enhancement of their neutron-capture
elements (the CEMP-no stars, according to Beers \& Christlieb 2005),
suggesting the existence of other possible origins for their carbon
excesses \citep[e.g., ][]{norris97b, aoki02a}.  \citet{aoki07} have
shown that the CEMP-no stars generally occur at very low [Fe/H];
extreme examples of this class of stars include HE~0107--5240 and
HE~1327--2326, two hyper metal-poor (HMP) stars with [Fe/H] below
$-5.0$ \citep{christlieb02, frebel05} and having very large carbon
excesses ([C/Fe] $\sim +4$), as well as the recently identified ultra
metal-poor (UMP; [Fe/H] = $-4.8$) star HE~0557-4840, with [C/Fe] $=
+1.6$ (Norris et al. 2007).

Among the CEMP stars for which chemical compositions have been
obtained from high-resolution spectroscopy, main-sequence turn-off
stars are expected to be of particular importance. In the case of mass
transfer in binary systems, the accreted material from the primary AGB
star has been mixed at least by the first dredge-up in red giants,
while turn-off stars might preserve pure material accreted from the
primary at their surfaces. In such cases, one can investigate the
efficiency of the production of carbon and neutron-capture elements in
AGB stars from abundance measurements of the secondary star. Another
interesting view arises from the suggested influence of so-called
thermohaline mixing (Charbonnel \& Zahn 2007; Stancliffe et al. 2007;
Denissenkov \& Pinsonneault 2007), which provides the possibility of
mixing the accreted surface material while the observed star is still
on the main-sequence or only slightly evolved, prior to first
dredge-up.  In this scenario, the contrast of the observed surface
abundances of CEMP turn-off stars with more evolved CEMP stars also
provides valuable clues to the nature of this proposed extensive
mixing process.

Very large new samples of CEMP stars have recently become available,
discovered during the course of the Sloan Digital Sky Survey (SDSS;
York et al. 2000; Adelman-McCarthy et al. 2007). Although originally
designed as an extragalactic survey, SDSS has also discovered large
numbers of VMP stars \citep{beers06}.  Although some of the CEMP stars
are the result of directed studies (Margon et al. 2002; Downes et
al. 2004), many of them have appeared among the calibration objects
used by SDSS to obtain spectrophotometric and telluric
corrections for other spectroscopic data.  These calibration stars
are primarily brighter, metal-poor main-sequence turn-off F- and
G-type stars. The ongoing extension to SDSS, SDSS-II (which includes
the program SEGUE: Sloan Extension for Galactic Understanding and
Exploration), is expected to provide tens of thousands of additional
VMP stars, at least several thousand of which are expected to be CEMP
stars. This paper reports the first application of abundance
measurements obtained with high-resolution spectroscopy for CEMP star
candidates found by the SDSS/SEGUE surveys.

In \S 2 we discuss the identification of our sample stars and the
observations that were carried out. A description of our analysis
techniques and results is provided in \S 3. In \S 4 we present a
discussion of our findings. The interesting kinematics of the SDSS/SEGUE CEMP
turn-off stars are discussed in \S 5.  We conclude with a few remarks in \S 6.

\section{Sample Selection and Observations}\label{sec:obs}

The Sloan Digital Sky Survey uses a CCD camera (Gunn et al. 1998) on a dedicated
2.5m telescope (Gunn et al. 2006) at Apache Point Observatory, New Mexico, to
obtain images in five broad optical bands ($ugriz$; Fukugita et al.~1996) over
approximately 10,000~deg$^2$ of the high Galactic latitude sky. The survey
data-processing software measures the properties of each detected object in the
imaging data in all five bands, and determines and applies both astrometric and
photometric calibrations (Pier et al. 2003; Lupton et al. 2001; Ivezi\'c et
al.~2004). Photometric calibration is provided by simultaneous observations with
a 20-inch telescope at the same site (Hogg et al.~2001; Smith et al.~2002;
Stoughton et al.~2002; Tucker et al.~2006).

\subsection{Sample selection and photometry data}

During the development of pipeline software for the determination of 
atmospheric parameters ({\teff}, {\logg}, [Fe/H]) for stars with available
photometry and spectroscopy from SDSS and SEGUE (the SEGUE Stellar Parameter
Pipeline; SSPP, see Lee et al. 2007a,b), it was noticed that a rather large
number of stars were present in the SDSS/SEGUE database with clearly enhanced CH
G-band strengths, and which were likely to be CEMP stars.  A list of over
1000 candidate CEMP stars was assembled, drawing in particular on the
calibration stars used by SDSS.  The sample formed the basis for a detailed
investigation of the frequency of CEMP stars in the SDSS database (see
Marsteller et al. 2006; Marsteller 2007).  

A handful of the brighter examples of the CEMP turn-off stars were identified for
carrying out a pilot study of their high-resolution spectroscopic abundances,
reported on herein. The imaging procedures used during the course of SDSS
are tuned for extragalactic observations. As a result, there exists a bright
limit corresponding to $g \sim 14.5$. Thus, the stars available for our study
are somewhat faint for high-resolution abundance analyses, even with 8~m-class
telescopes. 

\begin{figure} 
\includegraphics[width=12cm]{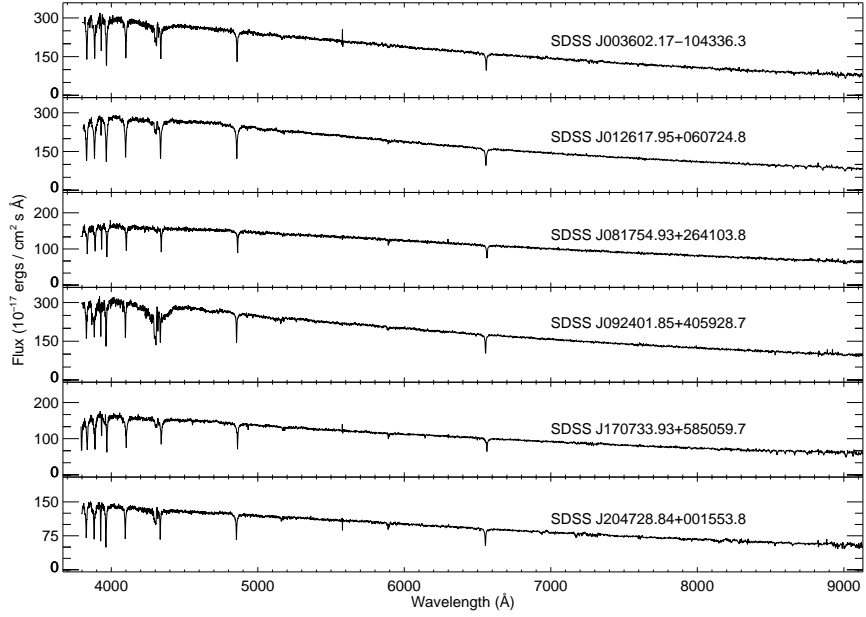} 
\caption[]{Medium-resolution, flux-calibrated SDSS spectra of our SDSS/SEGUE targets.}
\label{fig:sdss1} 
\end{figure}

\begin{figure} 
\includegraphics[width=12cm]{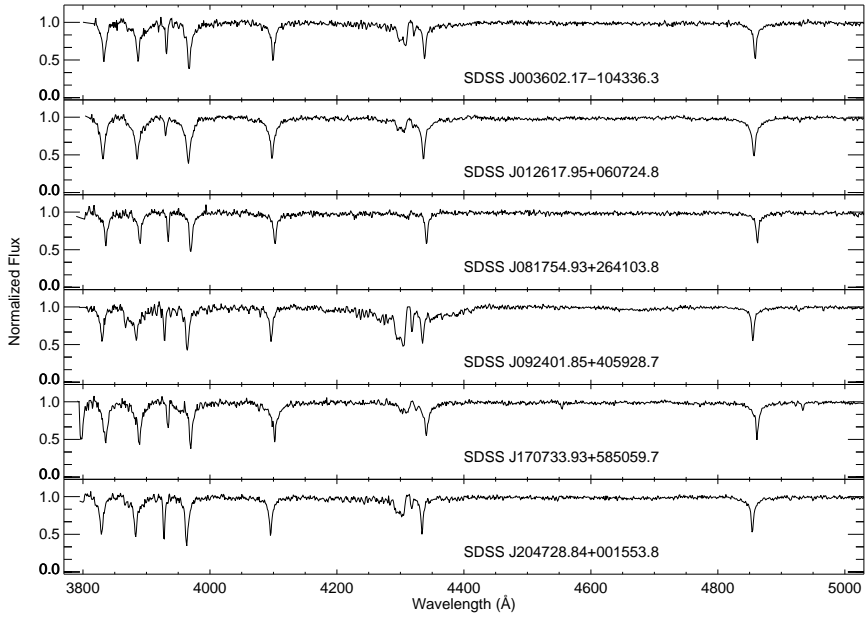} 
\caption[]{The blue range of the normalized SDSS spectra.}
\label{fig:sdss2} 
\end{figure}

The targets for the present observing program are listed in
Table~\ref{tab:obs}. Figures~\ref{fig:sdss1} and \ref{fig:sdss2}
show the medium-resolution ($R=\lambda/\delta \lambda \sim 2000$) SDSS spectra
of the targets. In addition to our primary objects, we selected two well-known
CEMP turn-off stars, LP~706--7 \citep{norris97a} and CS~29526--110
\citep{aoki02c}, as comparison stars.

The effective temperatures are primarily estimated from adopted $(V-K)_{0}$
colors (see \S~\ref{sec:ana}). The photometric data and reddening corrections
used in this work are listed in Table~\ref{tab:photo}. For the SDSS/SEGUE stars
the optical photometry information ($B$ and $V$) are obtained from the SDSS photometric
system, employing the following empirical transformations, obtained by
comparison with existing photometry for HK survey stars and subsequently
observed by SDSS (Zhao \& Newberg 2006):

$V = g - 0.561(g-r) - 0.004 $ \\

$B = g + 0.348(g-r) + 0.175$ \\

\noindent The photometric data for the comparison
stars are taken from \citet{beers07}.  The $K$ photometry
is obtained from the 2MASS catalogue \citep{skrutskie06}. The
interstellar reddening is estimated from the dust map of
\citet{schlegel98}; the extinction in the $V$ and $K$ bands is
obtained from the reddening relation provided in their Table 6.

\subsection{High-resolution spectroscopy}

High-resolution spectroscopy was obtained with the Subaru Telescope
High Dispersion Spectrograph (HDS; Noguchi et al. 2002) in September
2006 and February 2007. Our spectra cover the wavelength range from
4050 to 6800~{\AA}, with a gap between 5350 and 5450~{\AA} due to the
separation between the two detectors. A two-by-two pixel on-chip
binning procedure was applied. The resolving power of the spectra
obtained in 2006 is $R = 60,000$ (using a slit width of 0.6\arcsec),
while that obtained during the 2007 run is somewhat lower ($R =
45,000$) because a wider slit width (0.9\arcsec) was applied in order
to collect sufficient photons under relatively poor seeing
conditions. The total exposure times are listed in the third column
of Table \ref{tab:obs}. It was immediately obvious, during the course of the
observing run, that {\objf} exhibited a rapid variation in its radial velocity.
The exposure times for individual exposures for this object are listed
separately in Table~\ref{tab:s0353}. The total exposure time for this object in
Table~\ref{tab:obs} is the value for the spectrum used in the abundance analysis
(see \S~\ref{sec:ana}).

\begin{figure} 
\includegraphics[width=8cm]{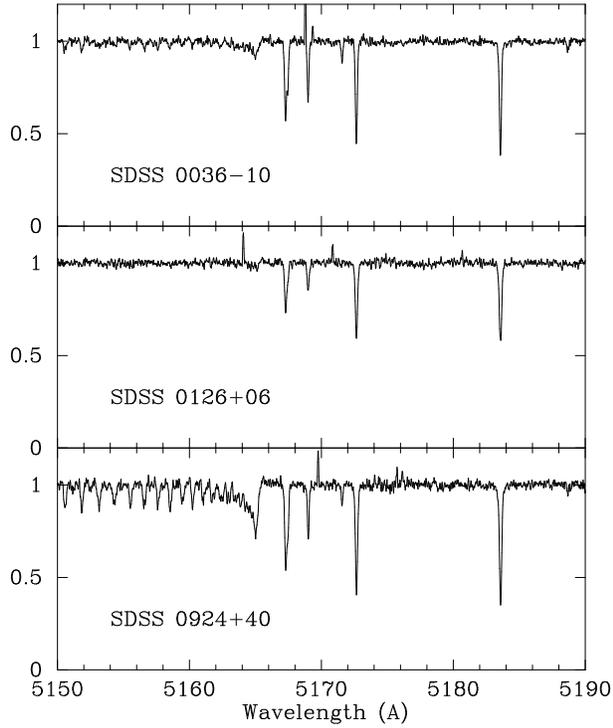} 
\caption[]{Examples of the Subaru spectra for the range including the C$_{2}$ Swan 0-0 band and the \ion{Mg}{1} triplet.}
\label{fig:sp} 
\end{figure}

Data reduction was carried out using standard procedures within
IRAF\footnote{IRAF is distributed by the National Optical Astronomy
Observatories, which is operated by the Association of Universities
for Research in Astronomy, Inc. under cooperative agreement with the
National Science Foundation.}.  Cosmic-ray hits were removed using the
procedure described by \citet{aoki05}. The wavelength scale was
determined using Th-Ar arc spectra obtained during the observing
nights. Examples of the spectra around 5170~{\AA} are shown in
Figure~\ref{fig:sp}. The photon counts per pixel (0.031~{\AA})
collected at 5100~{\AA} in the final spectra are listed in
Table~\ref{tab:obs}.

\subsection{Equivalent widths}\label{sec:ew}

The equivalent widths of atomic lines are measured by fitting Gaussian profiles
to (apparently) isolated absorption features. The list of atomic lines was made
using those of our recent studies for CEMP stars
\citep{aoki07}; the values are given in Table~\ref{tab:ew}. 


Equivalent widths of the interstellar \ion{Na}{1} D1 line ($\lambda 5990$) are
measured from a direct integration of the absorption features. The values are
given in the last line of Table~\ref{tab:ew}. We estimated $E(B-V)$ from the
\ion{Na}{1} absorption by applying the correlation found by \citet{munari97}.
The values are given in Table~\ref{tab:photo}. The agreement between the two
estimates of $E(B-V)$ is good in general. An exception is {\objc}, for which a
significantly larger $E(B-V)$ is obtained from the
\ion{Na}{1} line than from the dust map of \citet{schlegel98}. This is
discussed in \S \ref{sec:sdss}.

\subsection{Radial velocities}

Radial velocities for our program stars are measured from the
wavelengths of clean Fe lines; results are given in
Table~\ref{tab:obs}. The random errors of the measurements are
estimated from $\sigma_{v} N^{-1/2}$, where $\sigma_{v}$ is the
standard deviation of the values from individual lines and $N$ is the
number of lines used for the measurement. Note that the reported error
for the {\objd} value is substantially larger than for the rest of our
targets, because of the low $S/N$ ratio of the spectrum and the small
number of Fe lines used (5 lines).

\begin{figure} 
\includegraphics[width=8cm]{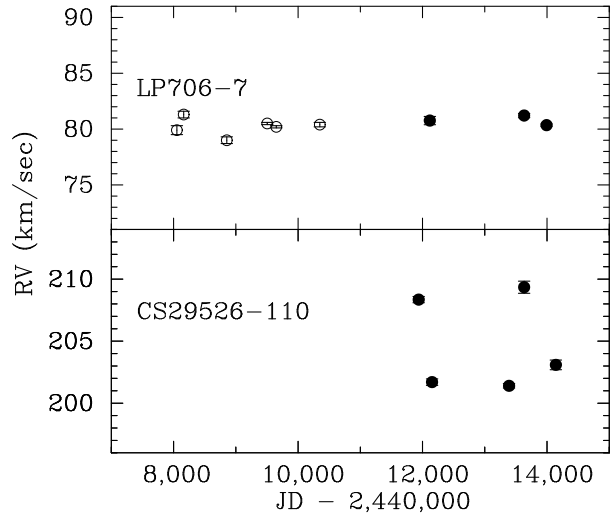} 
\caption[]{Radial velocity as a function of Julian Day number for {\LP} and
{\CS}. The filled circles indicate our measurements with the Subaru Telescope,
including the results by \citet{aoki02b}, while open circles mean the results by
\citet{norris97a} for {\LP} and by
\citet{aoki02c} obtained with the William Herschel Telescope for {\CS}.}
\label{fig:rv} 
\end{figure}

Figure~\ref{fig:rv} shows the radial velocities of {\LP} and {\CS} obtained by
the present work, along with additional measurements obtained during recent
service observing runs with the Subaru Telescope. For {\LP}, the results of
\citet{norris97a} are also shown. A clear variation of radial velocity is found
for {\CS}, indicating that this object belongs to a binary system, although the
orbital period is still unclear. On the other hand, no evidence of variability
is obtained for {\LP}, as has been discussed by \citet{norris97a}, even though
we have now obtained data extending over a range of some 6000 days.

\begin{figure} 
\includegraphics[width=8cm]{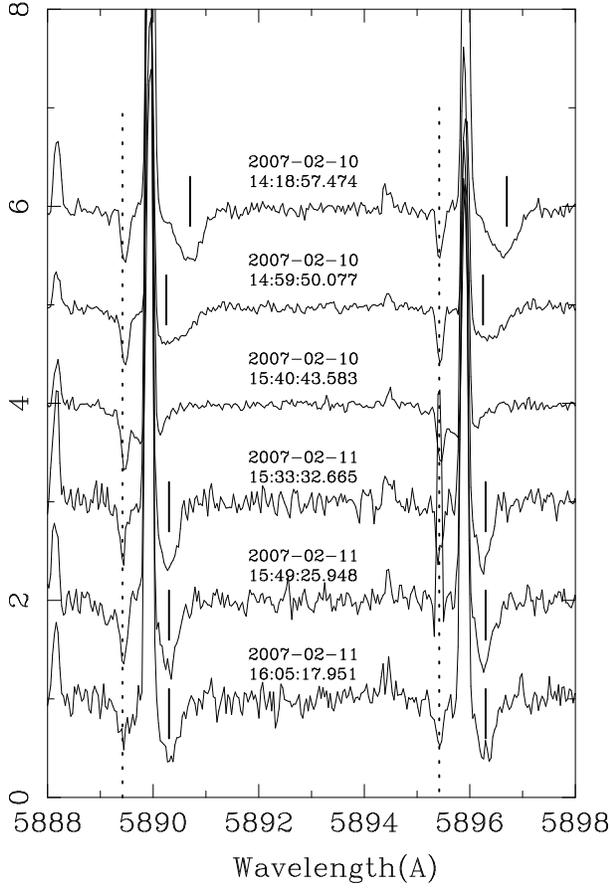} 
\caption[]{Spectra of the \ion{Na}{1} D lines region for {\objf}
  obtained by individual HDS exposures. The center of the exposure
  (UT) is presented for each spectrum. The exposure time is different
  between the two observing nights (see Table \ref{tab:s0353}). The
  positions of intersteller \ion{Na}{1} lines are shown by dotted
  lines, while the line positions of the stellar absorption are shown
  by solid lines. The emission from the Earth's atmosphere and
  interstellar absorption show no variation in wavelengths, while the
  stellar absorption lines show rapid changes in the 10 February (UT)
  spectra.}
\label{fig:s0353} 
\end{figure}

It was immediately noticed during the first observing night of 2007
that significant variations existed in the observed Doppler shift of
{\objf}. Figure \ref{fig:s0353} shows the spectra of this object
before any shifts have been obtained prior to co-addition for later
analysis. No velocity variation is found for the \ion{Na}{1} D
emission lines associated with the Earth's atmosphere at 5989.9 and
5995.9~{\AA}, nor for the interstellar absorption feature due to Na,
found about 0.5~{\AA} blueward of these emission lines, indicating
that the wavelength calibration was correctly carried out over all
exposures. In contrast, absorption due to the stellar \ion{Na}{1}
features quickly shifted in the spectra obtained in the first
observing night (the upper three spectra in the Figure), while no
significant variation is found for those obtained during the second
night. We note that the exposure times applied to the first and second
observing nights are 40 and 20 minutes, respectively. The shift in the
first night is about 18{\kms} per 40 minutes. Hence, the broad
absorption features of the first three spectra are almost certainly
due to a radial velocity shift during the exposures rather than
stellar rotation or macro-turbulence.
 
The radial velocities and line widths measured for each spectrum of {\objf} are
listed in Table~\ref{tab:s0353}. The measurements are made for strong
\ion{Na}{1}, \ion{Mg}{1}, and \ion{Ba}{1} lines, because measurements from the
\ion{Fe}{1} lines for individual exposures were quite uncertain. During the
course of this procedure, it was decided to exclude the third spectrum obtained
during the second night because of significant contamination from the twilight
sky. The UT and JD (heliocentric Julian day) of the central time of each
exposure is listed in the Table. A spectrum of this object, suitable for carrying
out the abundance analysis described below, is obtained by combining individual
spectra obtained during the first night, after applying the appropriate Doppler
corrections.

Radial velocities are also measured from the SDSS spectra, as listed
in Table~\ref{tab:rv_sdss}. The radial velocities of {\obja}, {\objc},
and {\obje} agree with the results from the Subaru spectra within the
measurement errors.  The radial velocity of {\objb} on JD = 2,452,932
is significantly higher than the other three measurements, suggesting
a radial velocity variation and binarity of this object. We note that
this object shows large over-abundances of carbon and neutron-capture
elements (\S \ref{sec:ana}) that are expected from mass transfer in a
binary system. The radial velocity of {\objf} from the SDSS spectrum
is within the variation found in the Subaru spectra given in
Table~\ref{tab:s0353}. Finally, the radial velocity of {\objd} from
the SDSS spectrum, 45.8 $\pm 3.5$ km sec$^{-1}$, is much higher than
the value obtained from the HDS spectrum. However, the measurement is only
once for each instrument, so further measurement is required to
derive any firm conclusions on the binarity of this object.

\section{Chemical Abundance Analysis and Results}\label{sec:ana}

\subsection{Stellar parameters}\label{sec:param}

We determine the effective temperatures from the $(V-K)_{0}$ colors
using the scale of \citet{alonso96}; these are listed in
Table~\ref{tab:photo} as {\teff}$(V-K)$. The $(V-K)_{0}$ values of
{\CS}, {\obja}, {\objb}, {\objc} and {\objf} are slightly lower than
the range for which the Alonso et al. scale (formula) is applicable
($V-K < 1.1$ for [Fe/H] $ < -1.5$). For these objects, we directly
estimate the effective temperature from Figure 8 of \citet{alonso96},
in which the correlation between $V-K$ and effective temperatures for
their calibration stars is shown.

The effective temperatures obtained from the $(B-V)_{0}$ colors using
the Alonso et al. scale are also listed in Table~\ref{tab:photo}.  For
CEMP stars, this color is sometimes severely affected by the presence
of molecular absorption, and it is not preferable for temperature
estimates. However, the molecular features of warm CEMP stars studied
here are not as significant as those for cooler stars, and would be
expected to have less of an affect on the observed colors. Moreover,
the errors in the $K$-band photometry for some of the fainter objects
in our sample are large (see below), and the $V-K$ colors are
relatively sensitive to the reddening correction. Hence, the effective
temperatures obtained from the $(B-V)_{0}$ colors are useful for
comparison purposes. The {\teff}$(B-V)$ of {\obja} and {\objf} agree
well with their {\teff}$(V-K)$ determinations. The {\teff}$(B-V)$ of
the two coolest stars in our sample, {\LP} and {\obje}, are lower than
their {\teff}$(V-K)$, perhaps as the result of their relatively strong
CH molecular bands affecting the $B$-band measurement. A similar
discrepancy between the two {\teff} estimates is found for {\objd},
even though this object exhibits no or perhaps only a modest ([C/Fe]$
\sim +1$) carbon overabundance. However, the error in the $K$
photometry for this star is relatively large, which might explain the
discrepancy.
 
The $T_{\rm eff}(V-K)$ of {\objb} is quite high (6800~K) for a VMP
turn-off star. However, the error of the $K$ photometry (0.12
magnitudes) and the reddening correction for this star
($E(B-V)=0.088$) are the largest among our sample. Moreover, the
$T_{\rm eff}(B-V)$ is about 200~K lower than the $T_{\rm
eff}(V-K)$. For this object we adopt {\teff} = 6600~K, which is
slightly lower than the estimate from the $V-K$ color.

The {\teff} of {\CS} is also quite high (6800~K). However, the
reported error of the $K$ photometry for this object is 0.03
magnitudes, and the reddening correction adopted (0.033 magnitudes) is
not large. For this object, the $R$ and $I$ photometry data are also
available \citep{beers07}. The {\teff} from $V-R$ and $V-I$ estimated
using the figures of \citet{alonso96} are 6700--6800~K, while the
{\teff}$(B-V)$ is 6500~K. We adopt the {\teff} from $(V-K)_{0}$ with
no modification for this object.

We now estimate the uncertainty in the adopted {\teff} for our program stars,
taking the error in the $(V-K)_{0}$ colors and the error in the scale of
\citet{alonso96} into consideration. The error of the $K$ photometry is the
dominant source of the uncertainty in the $(V-K)_{0}$ values for most objects.
The sensitivity of {\teff} to the color is approximately 150~K per 0.1 magnitude
in $V-K$. We adopt 100~K as the errors arising from the temperature scale for
stars with $(V-K)_{0}\geq 1.1$, for which Alonso et al.'s formula is applicable,
and 150~K for other objects, respectively. The uncertainties are 100-150~K for
relatively cool or bright objects ({\objd}, {\obje}, {\LP}, and {\CS}), and
150-200~K for others. The adopted errors of {\teff} in this study are listed in
Table~\ref{tab:param}.

The surface gravity, metallicity, and micro-turbulence for our program
stars are determined from an analysis of the \ion{Fe}{1} and
\ion{Fe}{2} lines, using the model atmospheres of
\citet{kurucz93}. The micro-turbulence ($v_{\rm turb}$) and gravity
(log $g$) are determined so that the derived Fe abundance is not
dependent on the strengths of Fe {\small I} lines, nor on the
ionization stages, respectively. An exception is {\objd}, for which
the number of useful \ion{Fe}{1} lines is too small to estimate the
micro-turbulence, and no \ion{Fe}{2} line is available to estimate the
gravity. We adopted typical values ({\logg} = 4.0 and {\vt} =
1.5~{\kms}) found for turn-off stars\footnote{If a lower gravity
($\log g =2.0$) is adopted for the case of a horizontal branch star,
the derived iron abundance is only slightly higher, while the derived
Sr and Ba abundances (see \S3.3) are about 0.6~dex lower.}. We note that this
object is excluded in the discussion of CEMP stars because its carbon
abundance is not determined by our analysis of the HDS spectrum (see
below); we only obtain a weak upper limit for [C/Fe]. The number of
\ion{Fe}{1} lines used in the analysis of {\objf} is also quite small,
due to the rapid changes of the radial velocity (see below). For this
object, {\vt}=1.5~{\kms} is also adopted. For {\objc}, a correlation
between the \ion{Fe}{1} line strengths and the derived Fe abundances
is found even if {\vt}$>2.0$~{\kms} is assumed. Since such a high
value of {\vt} is not known in turn-off stars, we adopt
{\vt}$=2.0$~{\kms} for this object. Larger errors in the gravity
($\sigma$[{\logg}]) and the micro-turbulence ($\sigma$[{\vt}]) are
adopted for these objects. The atmospheric parameters adopted in the
following abundance analyses and their corresponding errors are listed
in Table~\ref{tab:param}.

\begin{figure} 
\includegraphics[width=8cm]{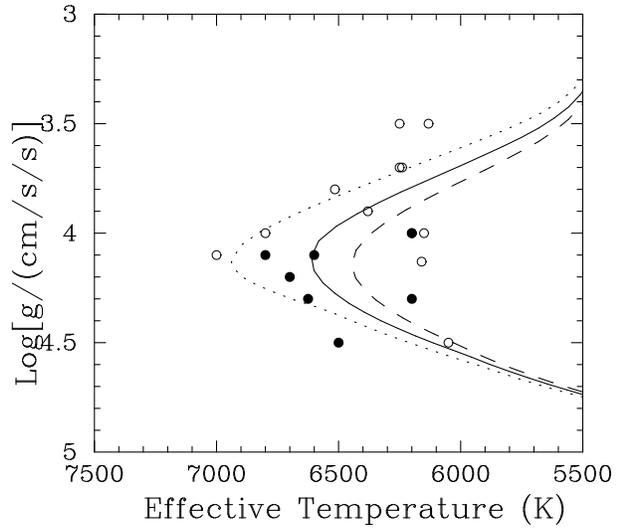} 
\caption[]{Surface gravity as a function of effective temperature for
  our sample (filled circles) and other CEMP turn-off stars from the
  literature given in Table~\ref{tab:cemp_to} (open circles). The
  dotted, solid, and dashed lines indicate the isochrones of
  \citet{y2} for [Fe/H] $=-2.5$, with ages of 10, 12, and 14 Gyr,
  respectively.}
\label{fig:teffg}

\end{figure}

Figure~\ref{fig:teffg} shows the estimated effective temperatures and
surface gravities for our sample (filled circles) along with other
CEMP stars studied in previous work (open circles; see below). The
lines are the isochrones by \citet{y2} for [Fe/H]$=-2.5$ and assumed
ages of 10, 12, and 14 Gyrs. Inspection of this Figure shows that our
objects fall around the turn-off region for old metal-poor stars,
although it is difficult to distinguish whether they are main-sequence
stars or subgiants. We note that if we adopt a higher effective
temperature (6800~K from $V-K$) for {\objb}, the surface gravity also
becomes quite high ({\logg}=4.9), far above the expected value based
on isochrones of VMP turn-off stars.

\subsection{Carbon abundance}

The carbon abundance estimates for our program stars are determined
from spectrum synthesis of the CH 4323~{\AA} band, as previously
described by \citet{aoki07}.  The sources of molecular data are
reported by \citet{aoki02c}. The oxygen abundance of [O/Fe]=+0.5 is
assumed in the analysis. We confirmed that the effect of assumed
oxygen abundance on the derived carbon abundances is negligible for
the range 0$<$[O/Fe]$<+2$ for a star with {\teff}$>6000$~K, in which
the fraction of carbon bound in the CO molecule is very small.

No signature of the CH band is detected in the HDS spectrum of {\objd},
so only an upper limit is determined (note that in the
medium-resolution SDSS spectrum, there is sufficient strength in this
band, and others, to obtain a detection, [C/Fe]$ = +1.2$; see \S
\ref{sec:sdss}). The determination of carbon abundance for {\objf},
based on the high-resolution spectrum alone, is very uncertain because
of the relatively low S/N ratio of the spectrum. The full set of
results is listed in Table~\ref{tab:abund}.

Carbon abundances are also measured from the C$_{2}$ Swan band at
5165~{\AA} for {\LP}, {\obja}, and {\obje}. The result for {\obje}
agrees well with that obtained from the CH band, while the carbon
abundances of {\LP} and {\obja} from the C$_{2}$ band are slightly
(0.1--0.2~dex) higher than those from the CH band, as was also found
for the CEMP subgiant LP~625--44 by \citet{aoki02b}. Although there
may exist a small systematic error in the determination of carbon
abundances from the C$_{2}$ band and/or from the CH band, the
measurements from the C$_{2}$ band confirm the reliability of carbon
abundance determination from the other molecular band.

\subsection{Abundances of other elements}

The abundances for most of the other elements are determined by a
standard analysis based on measured equivalent widths. The effects of
hyperfine splitting and isotope shifts are included in the analysis,
using \citet{mcwilliam98} for Ba, \citet{lawler01} for La, and
\citet{simons89} for Pb. Solar isotope ratios are assumed for Pb. For
Ba, we first measured the abundances neglecting the effect of
hyperfine splitting, and then applied the isotope ratios of the
r-process component in Solar System material for the two stars having
[Ba/Fe] $< 1$ ({\obja} and {\objd}), and s-process ratios for the
stars that exhibit Ba excesses, as was done previously by
\citet{aoki07}.

While Sr and Ba abundances are measured for all objects in our sample,
other neutron-capture elements are detected in only a few stars. The
abundances of Pb, which is a key element for investigation of
neutron-capture nucleosynthesis, are measured for {\LP}, {\CS},
{\objc} and {\obje}, while an upper limit is estimated for other
stars. The upper limit on the Pb abundance is calculated based on the
3$\sigma$ error of the equivalent-width measurement, estimated by
$\sigma_{W} = (\lambda n_{\rm pix}^{-1/2})/$($R$[S/N]), where $R$ is
the resolving power and $n_{\rm pix}$ is the number of pixels for
which equivalent width measurements are carried out
\citep{norris01}. The results are listed in Table~\ref{tab:abund}.

Six of our program stars exhibit large Ba over-abundances. In
particular, the over-abundance found for {\objf} is quite striking
([Ba/Fe] = +3.4). This star also exhibits a large excess of Sr
([Sr/Fe] = +2.25). Although the carbon abundance estimated from the CH
band for this object is very uncertain, we include this object in our
discussion of CEMP stars as an example of a star that is likely
affected by AGB nucleosynthesis (see \S \ref{sec:disc}).

\subsection{Uncertainties}

Random errors in our analysis, which include the uncertainty of the
equivalent-width measurements and in the adopted transition probabilities, are
estimated to be $\sigma N^{-1/2}$, where $\sigma$ is the standard
deviation of derived abundances from individual lines and $N$ is the
number of lines used in the analysis. When the number of lines are
smaller than four, the $\sigma$ of \ion{Fe}{1} ($\sigma_{\rm Fe}$) is
adopted in the estimates. Typical random errors are 0.05--0.15~dex,
depending on the number of lines used in the analysis.

We also estimate the error due to the uncertainty in equivalent-width
measurements for the \ion{Fe}{1} lines of {\obja}. A typical error in
equivalent width ($\sigma_{W}$) is estimated from the above
formula. The typical value for the \ion{Fe}{1} lines of {\obja} is
obtained to be 3~m{\AA}, assuming $\lambda = 5000$~{\AA} and $S/N
=70$. We added this value to the measured equivalent widths and
calculated the Fe abundance using the same model atmosphere as used in
the analysis. The derived iron abundance is 0.10~dex higher than the
original result. This value is comparable with the $\sigma_{\rm Fe}$
of 0.12~dex obtained for {\obja}. This result confirms that the random
errors of the abundance measurements are primarily due to the
uncertainties in the equivalent width measurements reflecting the
quality of the spectrum, although the $\sigma_{\rm Fe}$ also includes
the errors in the continuum placement and uncertainties of $gf$
values.

The errors due to the uncertainty of the atmospheric parameters are
estimated for {\LP} and {\CS}. Table~\ref{tab:err} lists the
sensitivity of the derived abundances ($\log \epsilon$ values) to the
changes of parameters. For other objects in our program, the
uncertainties are estimated by applying the data for the star of this
pair with the closest atmospheric parameters to the object under
consideration. Total uncertainties are obtained by adding these
values, in quadrature, to the random errors, and are listed in
Table~\ref{tab:abund}.

The chemical abundances of {\LP} are also determined using the updated
(NEWODF) ATLAS grid \citep{castelli03}, and the differences from those
based on the \citet{kurucz93} model are given in Table~\ref{tab:err}
($\Delta_{\rm ATLAS}$). The abundances using the NEWODF model are
lower by 0.05--0.14~dex. The effect of the difference of model
atmospheres on the derived abundances is systematic, and that on the
abundance ratios is not significant. We also applied the model
including the excesses of $\alpha$ elements, and confirmed the effect
on the derived abundances is smaller than 0.01~dex.

Further systematic errors could exist in our LTE analysis based on
one-dimensional (1D) model atmospheres. The non-LTE correction for Fe
abundances derived from the \ion{Fe}{1} lines might be the order of
+0.2~dex \citep[][ and references therein]{collet05, asplund05a},
although the values estimated are different between authors. The
direction of the correction for the 3D effect is opposite
\citep{asplund05a}. The most significant 3D effect would appear in the
carbon abundances determined from CH molecular features, that could
reach to $-0.7$~dex in the most metal-poor cases \citep{collet06}. In
order to obtain the corrections for these effect, 3D analyses based on
non-LTE calculation are required.

\subsection{Comparison with previous studies}

The elemental abundances of {\LP} (= CS~31062--012) and {\CS} were
studied by \citet{aoki02b} and \citet{aoki02c}. The atmospheric
parameters adopted by them for {\LP} are {\teff} = 6250~K, {\logg} =
4.5, [Fe/H] = $-2.55$, and {\vt} = 1.5 {\kms}, which are quite similar
to those in the present study.  Although the previous studies are
based on a spectrum covering only a blue range of wavelengths, and the
spectral line set used in the previous analyses is different from that
in the present study, the derived abundances of most elements agree
within 0.1~dex. The Cr abundance shows the largest discrepancy, on the
order of 0.17~dex, which is still within the 2~$\sigma$ range of the
measurement errors.

The effective temperature of {\CS} adopted by \citet{aoki02c} is 300~K
lower than that of the present analysis. The discrepancy of [Fe/H]
between the two measurements (0.32~dex) is well explained by the
difference in the adopted effective temperatures. The [C/Fe] values, after
correction for the difference in effective temperatures, also
agree within the measurement errors. The abundance ratios of other
elements ([X/Fe]) are relatively insensitive to the effective
temperature (see Table~\ref{tab:err}). It is clear that the results
for Cr and Ni from the two studies exhibit significant
discrepancies. However, the measurements for these elements are based
on only one line for each; the results might not be expected to be as
reliable as those for other elements. The [Ba/Fe] derived from the
present work is 0.28~dex higher than that of \citet{aoki02c}. This
result is not explained by the differences of adopted atmospheric
parameters. While the previous measurement is based on only the two
very strong resonance lines, our present analysis added two red lines
which are suitable for abundance determination, so the new measurement
should be more reliable than the previous one.

\subsection{Comparison with the estimates from SDSS/SEGUE spectra}\label{sec:sdss}

Table ~\ref{tab:comp} provides a comparison of the atmospheric
parameters and carbon abundances between the estimates obtained from
the SDSS spectra and the present measurements.

For the SDSS spectra, we begin by adopting the stellar parameters
obtained by the SSPP (Lee et al. 2007a,b). Based on these, we generate
synthetic spectra for each star in the region between
4200--4400~{\AA}. The model atmospheres used are the NEWODF models of
Castelli \& Kurucz (2003). The synthetic spectra are generated using
the turbospectrum synthesis code \citep{alvarez98}, which employs
line broadening according to the prescription of \citet{barklem98} and
\citet{barklem05}. The atomic line data are taken mainly from the VALD
compilation (as of 2002) (Kupka et al. 1999), and updated from the
literature, whenever possible. The molecular species CH and CN are
provided by B. Plez (Plez \& Cohen 2005). We adopted the solar
abundances by Asplund, Grevesse \& Sauval (2005a). The synthetic
spectra are generated with a initial resolving power $R= 10^{6}$, then
were smoothed to the SDSS resolution and rebinned to 1~{\AA} pixels.

We find the best match to the region around the G band (4323~{\AA} and
4325~{\AA}) by changing the carbon abundance of the synthetic spectra
in order to minimize the discrepancy with the observed spectra. We
estimate that the errors in the derived [C/Fe] arising from errors in
the stellar parameters from the SSPP is on the order of 0.35 dex.

The effective temperatures derived from the SDSS spectra agree with
the values adopted by the present work, based on colors, to within about
100~K. An exception is that for {\objc}, for which the SDSS estimate
is 370~K higher than the value adopted in the above analysis. For this
object, larger interstellar reddening is derived from the \ion{Na}{1}
absorption than from the dust map that is adopted in the analysis
(\S~\ref{sec:ew}). If the $E(B-V)$ from the \ion{Na}{1} measurement is
adopted, the {\teff} is as high as 7000~K, and agrees with the
estimate from the SDSS spectrum.

In contrast to the agreement of effective temperatures, {\logg} values
estimated from SDSS spectra are systematically lower than those
determined by our analyses from Subaru spectra. Although the results
for {\objc} from the two estimates appear to agree well, a similar
discrepancy would result if the same effective temperature is adopted
in the estimate of gravity. Although further study or calibration to
resolve the discrepancy is desired, it should be noted that the
{\logg} values from SDSS spectra are already useful to estimate the
evolutionary status of the targets (i.e., in order to distinguish
giants and main-sequence turn-off stars).

The abundance ratios [Fe/H] and [C/Fe] from the two measurements agree fairly
well, as found in Table~\ref{tab:comp}. The discrepancy in [Fe/H] for {\objc}
(0.4~dex) is well explained by the difference in the estimated {\teff}. No CH
absorption feature is detected in the Subaru spectrum of {\objd}. A very weak CH
band is found in the SDSS spectrum of this object (see Figure 2), and we have
derived [C/Fe] = +1.19 $\pm 0.35$, based on its strength. \citet{marsteller07}
estimated that the detection limit of the CH feature in SDSS spectra can be as high as
[C/H]$\sim -1$ for stars with {\teff} $\sim 6300$~K. Thus, {\objd} could also be
mis-identified as a carbon-enhanced object in our sample selection, which was
carried out before detailed investigations for SDSS sample were made. Further
calibrations, or strict estimates for the detection limits of CH absorption in
SDSS spectra as a function of {\teff}, are desired for more efficiently
selecting carbon-enhanced objects from SDSS/SEGUE spectra.

\section{Discussion}\label{sec:disc}

\subsection{Elemental abundances of CEMP turn-off stars}

The present analysis has obtained elemental abundances for five new
CEMP stars selected from the SDSS/SEGUE surveys, as well as for two
known CEMP stars. The remaining program star in our sample ({\objd})
is excluded from the following discussion, because no clear excess of
carbon nor neutron-capture elements has been found in the
high-resolution spectrum. All seven CEMP stars are classified as
main-sequence turn-off stars, with {\teff}$>6000$~K. In order to
better investigate the nature of such stars, we have compiled all
known CEMP stars from the literature having {\teff} higher than 6000~K
(Table~\ref{tab:cemp_to}). In this section,
we discuss the abundance distributions of carbon and the
neutron-capture elements for these CEMP turn-off stars.

\subsection{Distribution of C and Ba abundances}

\begin{figure} 
\includegraphics[width=8cm]{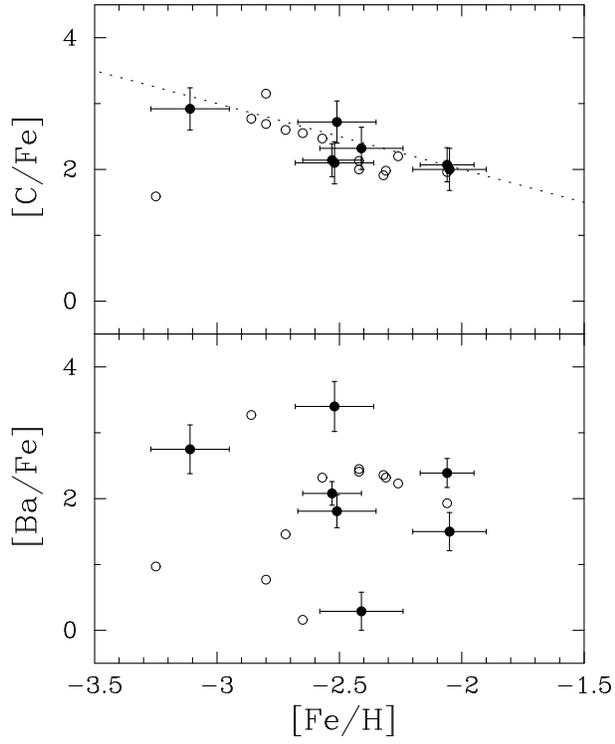} 
\caption[]{The abundance ratios of [C/Fe] and [Ba/Fe] as functions of
  [Fe/H]. Filled circles are objects studied by the present work,
  while open ones are from the literature given in
  Table~\ref{tab:cemp_to}. The dotted line in the upper panel means
  the line for [C/H]=0.}
\label{fig:cbafe} 
\end{figure}

Figure~\ref{fig:cbafe} shows the C and Ba abundance ratios as a
function of [Fe/H]. All objects but one exhibit very large excesses of C
([C/Fe] $\gtrsim +2$).  The exception is CS~29528--041, which has
[C/Fe]= +1.59 at [Fe/H]$=-3.25$ \citep{sivarani06}. Excluding this
object, a clear correlation is found between [C/Fe] and [Fe/H]. The
correlation indicates that there exists a constant [C/H] among these
stars.

\begin{figure} 
\includegraphics[width=8cm]{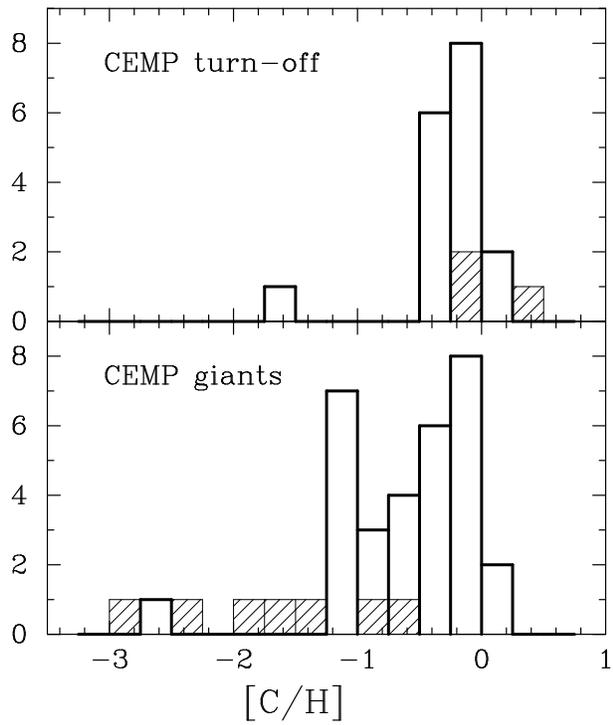} 
\caption[]{[C/H] distributions for CEMP turn-off stars (upper) and
  giants (lower). See text for the selection criteria for these
  samples. The open histogram with strong lines indicates the objects
  having excesses of Ba ([Ba/Fe] $> +0.5$). The hatched one is for
  Ba-normal stars.}
\label{fig:hist_ch} 
\end{figure}

Figure~\ref{fig:hist_ch} is a histogram of the [C/H] values for the
CEMP turn-off stars. The distribution is compared with that of the 31
CEMP giants selected from \citet{aoki07} that have $\log
L$/L$_{\odot}$ higher than 1.5, where $L$ is the luminosity estimated
from the stellar parameters assuming a constant stellar mass (see Aoki
et al. 2007). The Ba-enhanced (CEMP-s) stars are shown by open bars,
while Ba-normal (CEMP-no) stars are shown by the hatched bars.  A
glance at this figure shows that the [C/H] ratios of the CEMP-s
turn-off stars distribute within a narrow range around [C/H]$\sim 0$,
with the exception of the one object mentioned above (CS~29528--041).
The average and standard deviations of the [C/H] ratios for these two
CEMP-s samples are $<$[C/H]$>=-0.18$ and $\sigma$([C/H])$=0.18$ for
the turn-off stars, and $<$[C/H]$> = -0.54$ and $\sigma$([C/H])$=0.40$
for the giants. The two objects having the lowest [C/H] among turn-off
stars (CS~29528--041) and giants (CS~30322-023) are excluded in the
calculation of these statistics.

The lower portion of the [C/H] distribution might be affected by a
temperature-related selection bias for CEMP stars identified on the
basis of the CH molecular bands, which is significantly weaker in
turn-off stars than in giants. Our previous investigation for the
detection limit showed that the CH band of a turn-off star with
{\teff}$\sim 6400~K$ and [C/H]$\sim -1.6$ has depths of 2\%, which is
a conservative detection limit in high-resolution spectra. For the
carbon-enhanced stars selected from the lower-resolution spectra, such
as the SDSS/SEGUE sample, \citet{marsteller07} estimated that the
selection of CEMP stars from the CH band is complete for stars having
[C/H] $=-1$ and $=-0.3$ for stars with {\teff} = 6000 and 6500~K,
respectively. Hence, in order to investigate the complete distribution
of [C/H] for turn-off stars, abundance studies of candidate metal-poor
stars that are selected regardless of their CH band strengths in
medium-resolution spectra are required. Note, however, that the sample
of CEMP turn-off stars in Table~\ref{tab:cemp_to} includes stars with
{\teff} as low as 6000~K, and that several stars were observed on
programs that did not focus on carbon-enhanced stars
\citep[e.g. ][]{cohen04}. The absence of stars with $-1<$ [C/H] 
$<-0.5$ in the sample of CEMP turn-off stars suggests that such stars
are rare compared with CEMP stars with higher [C/H] values. We note
that we cannot derive any conclusion for the lower [C/H] range ([C/H]$
< -1.0$). There may well exist a number of CEMP turn-off stars that
have not yet been identified by the surveys to date. This range is
particularly important for studies of the Ba-normal CEMP stars, as
discussed by \citet{aoki07}.

\citet{aoki07} showed that the [C/H] distribution for 54 CEMP stars
with Ba excesses, including turn-off stars, subgiants, and giants,
exhibits a peak in the range $-0.5<$ [C/H] $<0.0$, and a cut-off at
[C/H] $\sim 0$. This was interpreted as an indication that (1) the
[C/H] ratios produced by AGB stars are almost constant at [C/H] $\sim
0$, independent of metallicity, and (2) the carbon-enhanced material
transferred from AGB stars to the companion is directly observed, or
is only slightly diluted through the evolution from turn-off stars to
giants. In this study we confirmed the absence of objects having [C/H]
$>>0$, found that the average of the [C/H] values for CEMP turn-off
stars is higher than that of giants, and that their dispersion is
smaller. This result supports the interpretation of \citet{aoki07}.

\citet{stancliffe07} investigated the process of thermohaline mixing
in main-sequence stars that received carbon-enhanced material from a
companion AGB star across a binary system. They predicted that the
accreted material quickly mixes throughout 90\% of the star, and that
the enhanced carbon is diluted in main-sequence stars as a result. The
C abundance is predicted to change only slightly after the receiving
star evolves through first dredge-up.  This is not supported by the
comparison of [C/H] distributions in Figure~\ref{fig:hist_ch},
although the possible bias in the sample selection could slightly
affect the comparison. Our present observational result suggests that
the [C/H] ratios measured for turn-off stars represents the values
produced by the donor AGB stars, and the surface carbon abundance
decreases in some CEMP stars during their evolution after the first
dredge-up \footnote{One possible interpretation is that the mass
accreted from AGB stars is much larger than that assumed in the models
of \citet{stancliffe07} in most cases, and the dilution in
main-sequence stars is not as significant as predicted in their
models.}. It should be noted that the [C/H] values found in CEMP
turn-off stars (i.e. [C/H]$\sim 0$) agree well with predictions from
AGB models \citep[e.g. ][]{vandenhoek97}, as discussed by
\citet{aoki07}. This agreement supports the above interpretation.  


The carbon excesses of CS~29528--041 and CS~30322--023 are
exceptionally small among the sample of CEMP turn-off stars and giant
stars, respectively\footnote{After this paper is submitted, an analysis
result for the double-lined spectroscopic binary CS~22964--161 was
reported by \citet{thompson08}. The both components are CEMP turn-off
stars showing large excesses of neutron-capture elements. The [C/H] of
this system is $-1.2$, and is another example of CEMP turn-off stars
having relatively low [C/H] values.}. It is noteworthy that these two
stars both exhibit very large excesses of nitrogen -- the [N/Fe] of
CS~29528--041 is +3.0 \citep{sivarani06} and that of CS~30322--023 is
+2.8 \citep{masseron06}. Their [(C+N)/H] values are $-0.9$ and $-1.3$,
respectively, which are not by far lower than those of other stars,
although the nitrogen abundances are not determined for several stars
including our SDSS sample. CS~30322--023 is a highly evolved giant,
and possibly presently in the AGB stage (according to Masseron et
al.); its surface composition could have been altered significantly
during its evolution.  The observed nitrogen excess of CS~29528--041
should instead be a direct result of the nucleosynthesis in a donor
AGB star. As discussed by \citet{sivarani06}, CS~29528--041 might have
been polluted by an intermediate-mass AGB star in which nitrogen is
enriched by the hot bottom burning process.


\subsection{Neutron-capture elements}

\begin{figure} 
\includegraphics[width=8cm]{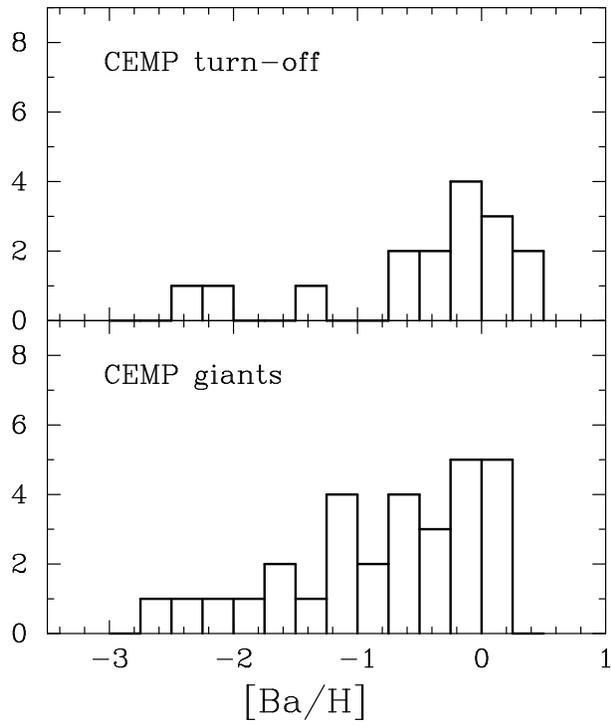} 
\caption[]{The same as Fig.~\ref{fig:hist_ch}, but for [Ba/H]. Only the
  histogram for Ba-enhanced ([Ba/Fe] $>+0.5$) stars are shown.}
\label{fig:hist_bah} 
\end{figure}

Based on their observed [C/H] distribution, we have interpreted the
surface abundances of CEMP turn-off stars to represent the yields of
the AGB donors. If this is indeed the case, [Ba/H] can be used as an
indicator of the s-process efficiency in the donor
stars. Figure~\ref{fig:hist_bah} depicts the distribution of [Ba/H]
for the CEMP turn-off stars and giants that show excesses of Ba
([Ba/Fe] $ > +0.5$). The [Ba/H] of the turn-off stars exhibits a wider
distribution than was observed for their [C/H]. The average and
standard deviation of the [Ba/H] for the 16 CEMP turn-off stars
discussed in the context of their [C/H] are $<$[Ba/H]$> = -0.25$ and
$\sigma$([Ba/H]) =0.68 dex, respectively. The wider distribution of
[Ba/H] of these stars, as compared to [C/H], implies that the Ba
abundances produced by AGB stars are likely to have a significantly
larger intrinsic dispersion. 

The abundance ratios currently observed are those of the material
transferred from the donor AGB stars, which would be dependent on the
evolutionary phase in which the mass transfer occurred. The abundances
of carbon and Ba are expected to increase with increasing the number
of thermal pulses. Detailed comparisons with model predictions for
surface abundances after each thermal pulse would be required. In
such comparisons, a strong constraint is the almost constant C/H
ratios found in most CEMP turn-off stars.

\begin{figure} 
\includegraphics[width=8cm]{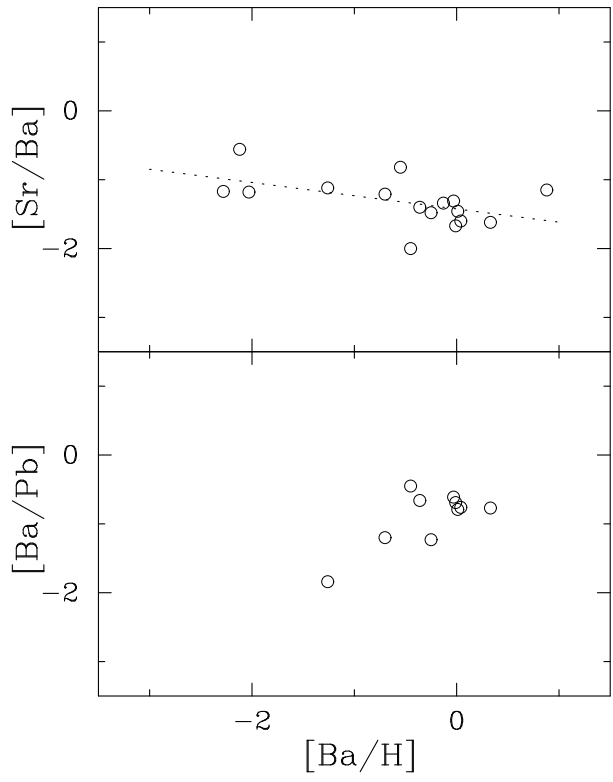} 
\caption[]{The abundance ratios of [Sr/Ba] (upper) and [Ba/Pb]
  (lower), as functions of [Ba/H], for CEMP turn-off stars.}
\label{fig:srbapb} 
\end{figure}

Figure \ref{fig:srbapb} shows the [Sr/Ba] and [Ba/Pb] ratios of CEMP
turn-off stars as a function of [Ba/H]. Here, the three elements Sr,
Ba, and Pb are regarded as representing the yields at the three
abundance peaks of the s-process, corresponding to the neutron magic
numbers 50, 82, and 126. The [Sr/Ba] ratios are distributed over a
narrow range (a standard deviation of 0.34~dex) around [Sr/Ba] $\sim
-1.3$. However, they also exhibit a statistically significant
anti-correlation with [Ba/H] -- the null hypothesis that no
correlation exists between [Sr/Ba] and [Ba/H] is rejected by a simple
t-test at the 98\% confidence level. That is, the production
efficiency of the heavy neutron-capture elements, with respect to the
light ones, apparently increases slightly with the total production
efficiency of neutron-capture elements.  It should be noted that the
[Sr/Ba] ratios in these stars are lower than the prediction from
models of the s-process in AGB stars by \citet{busso01}: the
[ls/hs]($=-$[hs/ls]) values of their calculations, where ls and hs
mean the elements in the first and second abundance peaks of the
s-process, range between 0 to $-1$, depending on stellar mass and
choice of $^{13}$C pocket, for the metallicity range of
[Fe/H]$<-2$. The observations suggest more efficient production of Ba
(hs), or less efficient production of Sr (ls), than model predictions.


A similar correlation was found from the abundance measurements for carbon-rich
post-AGB stars by \citet{reyniers04}. They reported a positive correlation
between the enhancement of s-process elements ([s/Fe], which is the mean of
the abundance ratios of several s-process elements) and the abundance ratios of
heavy to light neutron-capture elements ([hs/ls]). We note that the efficiency
of the neutron-capture nucleosynthesis is represented by [s/Fe] in
\citet{reyniers04}, while it is evaluated by [Ba/H] in our investigation.
However, a correlation between [Sr/Ba] and [Ba/Fe] is also found in our
sample at a similar confidence level as that between [Sr/Ba] and [Ba/H]. 




The lower panel of Figure \ref{fig:srbapb} suggests that there may
exist a correlation for [Ba/Pb] with [Ba/H], but this essentially
depends on only one object (HE~0024--2523), with [Ba/H] $=-1.36$ and
[Ba/Pb] $=-1.84$.  It should be kept in mind that the detection of Pb
lines is much more difficult than those of Sr and Ba, which could
result in a lack of objects having high [Ba/Pb] and low [Ba/H].  The
[Ba/Pb] ratios observed are higher than the prediction from models of
the s-process by \citet{busso99}, as has been argued by previous
studies \citep[e.g. ][]{cui06}. Further studies of Pb abundances, as well
as modeling of the s-process to explain the discrepancy between the
observations and predictions, are clearly required in order to better
understand the overall neutron-capture nucleosynthesis process in AGB
stars.

\subsection{{\objf}: an object belonging to a close binary?}

As discussed in \S~\ref{sec:obs}, {\objf} exhibited a rapid variation
of radial velocities on February 10, 2007, suggesting that this object
belongs to a close binary system. However, it is interesting that no
clear variation was found in the February 11 spectra, where the radial
velocity is close to the middle of the three spectra obtained on
February 10.  This result could well imply a large eccentricity of the
binary system, although the radial velocity measurements are still too
sparse to confidently derive orbital parameters. Further radial
velocity monitoring of this object is clearly desirable.

The observed excesses of neutron-capture elements in this object are
very large (e.g., [Ba/Fe] = +3.4). Indeed, the [Ba/H] value of this
object is the highest among the CEMP stars shown in
Figure~\ref{fig:hist_bah}. The [Sr/Ba] ratio ([Sr/Ba] $=-1.15$) is a
typical value found for Ba-enhanced CEMP stars
(Figure~\ref{fig:srbapb}), as compared to that of r-process-enhanced
stars ([Sr/Ba]$\sim -0.4$; e.g.  Sneden et al. 2003), suggesting that
the neutron-capture elements of this star originated from operation of
the s-process in an AGB donor star.

\citet{lucatello03} studied the CEMP turn-off star HE~0024--2523
([Fe/H]$=-2.7$), which they showed to be a short-period spectroscopic
binary.  This star has a orbital period of 3.4~days, with a very small
eccentricity (0.01). \citet{sivarani06} also reported a candidate CEMP
turn-off close binary, CS~22958--042, for which a significant radial
velocity change was found based on two exposures taken during a single
observing run. {\objf} is possibly another example of CEMP stars
belonging to close binary systems. \citet{lucatello03} proposed that
HE~0024--2523 underwent a past common-envelope phase with its
companion that has become an AGB star. A similar past history may be
applicable to {\objf}.

The stellar parameters of {\objf} ({\teff} = 6700~K and {\logg} = 4.2)
are almost the same as those of HE~0024--2523 ({\teff} = 6625~K and
{\logg} = 4.3).  On the other hand, the abundance ratios of the
neutron-capture elements are different between these two stars -- while
HE~0024-2523 exhibits a very large excess of Pb ([Pb/Fe] = +3.3) and
moderate over-abundances of Ba ([Ba/Fe] = +1.46), {\objf} has a very
large overabundance of Ba ([Ba/Fe] = +3.40). Our derived upper limit
on the Pb abundance ratio of {\objf} is still quite high ([Pb/Fe] $ <
+3.7$), so it does not constrain this comparison at present. We note
that the possible close binary CS~22958--042 exhibits no excess of
neutron-capture elements, but shows a large over-abundance of Na, as found
also for {\objf}. Thus, large variations of chemical abundances are found
even in these three (candidate) close binary systems.  Further
abundance measurements, as well as radial velocity monitoring, for
{\objf} are clearly desired to understand the evolution of such binary
systems.

\section{Kinematics}

Table~\ref{tab:kinematics} provides kinematic data for our SDSS
targets.  Distances (expected to be accurate to on the order of
10-15\%) for these stars are estimated using the methods described by
\citet{beers00}, under the assumption that they are main-sequence
stars, as suggested by the surface gravities determined in this
study. Proper motions are obtained based on the re-calibrated USNO-B2
catalog, as described by Munn et al. (2004), and are expected to be
accurate to on the order of 3 mas/year. The radial velocities are
taken from our high-resolution estimates listed in Table 1, with the
exception of SDSS 1707+58, where we adopt the value measured from the
SDSS spectrum. The space motions, errors in the space motions, and the
other derived quantities listed in Table~\ref{tab:kinematics} are
obtained following the procedures of \citet{beers00}.

All six of the SDSS CEMP stars exhibit a derived $r_{\rm max}$ (maximum distance
from the Galactic center achieved during the course of their orbits) in excess of
10 kpc from the Galactic center. Four of the stars exhibit $Z_{\rm max}$
(maximum height above or below the Galactic plane achieved during the course of
their orbits) values larger than 10~kpc, or significant retrograde motions,
indicating that these stars may belong to the outer-halo population of
our Galaxy, according to the criteria of \citet{carollo07}. This is also
reminiscent of the apparent excess of CEMP stars with increasing distance from
the plane reported by Frebel et al. (2006). Although the sample size is too
small to derive any firm conclusions, the fraction of outer-halo stars among the
CEMP stars appears to be quite high compared with the fraction found for
metal-poor stars ([Fe/H] $<-2.2$) studied by \citet{carollo07}. A comparison of
the fraction of CEMP stars associated with the inner- and outer-halo
populations, at a given metallicity, potentially constrains the initial mass
function of early-generation stars, as discussed recently by \citet{tumlinson07}
and
\citet{komiya07}, and is clearly of interest for additional study, in particular
given the very large samples of CEMP stars identified in SDSS/SEGUE.

\section{Concluding remarks}

Chemical compositions of seven CEMP turn-off stars are determined.
Six stars among them exhibit a large excess of Ba, signature of a
contribution by the nucleosynthesis in an AGB star. The distribution
of carbon abundances in these stars suggest that the surface of such
stars preserves the material transferred from the AGB star that was the
erstwhile primary star in a binary system. If this is the case, the
relatively wide distribution of Ba abundances ([Ba/H]) indicates a
diversity of the efficiency of the s-process in metal-poor AGB stars.
Further studies to identify the physical mechanism that produces such
diversity are clearly desired.

The present study is the first application of high-resolution
spectroscopy to candidate CEMP stars from the SDSS and SEGUE sample.
Comparisons of our results on stellar parameters and chemical
abundances with the estimates from the SDSS spectra confirmed that the
selection of metal-poor stars works well in general. The SDSS/SEGUE
survey is providing a large sample of candidate metal-poor stars.
High-resolution spectroscopy for such stars in the near future will reveal the
chemical abundance trends in the lowest metallicity range, as well as be useful
for exploring the possible dependence of their chemical properties on
their derived kinematics.

\acknowledgments

Funding for the SDSS and SDSS-II has been provided by the Alfred
P. Sloan Foundation, the Participating Institutions, the National
Science Foundation, the U.S. Department of Energy, the National
Aeronautics and Space Administration, the Japanese Monbukagakusho, the
Max Planck Society, and the Higher Education Funding Council for
England. The SDSS Web Site is http://www.sdss.org/.

The SDSS is managed by the Astrophysical Research Consortium for the
Participating Institutions. The Participating Institutions are the
American Museum of Natural History, Astrophysical Institute Potsdam,
University of Basel, University of Cambridge, Case Western Reserve
University, University of Chicago, Drexel University, Fermilab, the
Institute for Advanced Study, the Japan Participation Group, Johns
Hopkins University, the Joint Institute for Nuclear Astrophysics, the
Kavli Institute for Particle Astrophysics and Cosmology, the Korean
Scientist Group, the Chinese Academy of Sciences (LAMOST), Los Alamos
National Laboratory, the Max-Planck-Institute for Astronomy (MPIA),
the Max-Planck-Institute for Astrophysics (MPA), New Mexico State
University, Ohio State University, University of Pittsburgh,
University of Portsmouth, Princeton University, the United States
Naval Observatory, and the University of Washington.

W.~A. is supported by a Grant-in-Aid for Science Research from JSPS
(grant 18104003). T.~C.~B., B.~M., and T.~S. acknowledge support by
the US National Science Foundation under grants AST 04-06784 and AST
07-07776, as well as from grant PHY 02-16783; Physics Frontier
Center/Joint Institute for Nuclear Astrophysics (JINA). D.~C. is
grateful to JINA for support of her long-term visitor status at
Michigan State University, where the kinematical analysis took place.
J.E.N. acknowledges support from the Australian Research Council under
grant DP0663562.

\clearpage
\begin{deluxetable}{lccccc}
\tablewidth{0pt}
\tabletypesize{\footnotesize}
\tablecaption{\label{tab:obs} PROGRAM STARS AND OBSERVATIONS}
\tablehead{
Star    & IAU Name &  Exp.\tablenotemark{a} & Counts\tablenotemark{b} &  Obs. date (JD) & $V_{\rm helio}$ ({\kms}) ~  \\
}
\startdata
{\LP}     &                                 & 20  & 16500 & 14 Sep 2006 (2453992.87) & $  80.35 \pm 0.10$ \\
{\CS}   &                          & 40  & 8000  & 10 Feb 2007 (2454141.72) & $ 203.09 \pm 0.39$ \\
{\obja} & SDSS J003602.17-104336.3 & 120 & 5350  & 14 Sep 2006 (2453992.95) & $-146.18 \pm 0.18$ \\
{\objc} & SDSS J012617.95+060724.8 & 120 & 5150  & 14 Sep 2006 (2453993.04) & $-272.24 \pm 0.28$ \\
{\objd} & SDSS J081754.93+264103.8 & 94  & 1050 & 10 Feb 2007 (2454141.76) & $   1.74 \pm 2.52$ \\
{\obje} & SDSS J092401.85+405928.7 & 160 & 5250 & 10 Feb 2007 (2454141.87) & $-366.16 \pm 0.23$ \\
{\objf} & SDSS J170733.93+585059.7 & 117 & 2300 & 10 Feb 2007 (2454142.10) & ... \\
{\objb} & SDSS J204728.84+001553.8 & 160 & 3000  & 14 Sep 2006 (2453992.75) & $-417.92 \pm 0.20$ \\
\enddata
\tablenotetext{a}{Exposure time (minutes)}
\tablenotetext{b}{The photon counts per pixel (0.18{\kms}) at 5100~{\AA}}
\end{deluxetable}

\begin{deluxetable}{lcccccccccc}
\rotate
\tablewidth{0pt}
\tabletypesize{\footnotesize}
\tablecaption{PHOTOMETRY DATA \label{tab:photo}}
\tablehead{
object        &   $V$\tablenotemark{a} &$\sigma(V)$ & $B-V$\tablenotemark{a} & $\sigma(B-V)$ & $K$ & $\sigma(K)$ &$E(B-V)$\tablenotemark{b} & $E(B-V)$\tablenotemark{b} & {\teff}$(V-K)$ &  {\teff}$(B-V)$  
}
\startdata
{\LP}   &   12.098 &  0.001 & 0.467  & 0.003 & 10.771 &  0.017 & 0.022 & 0.007 & 6206 & 5933 \\ 
{\CS}   &   13.352 &  0.004 & 0.356  & 0.007 & 12.367 &  0.023 & 0.033 & 0.025 & 6800 & 6500 \\
{\obja} &   15.540 &  0.004 & 0.320  & 0.006 & 14.388 &  0.080 & 0.027 & 0.010 & 6500 & 6600 \\ 
{\objc} &   15.525 &  0.004 & ...    & ...   & 14.468 &  0.077 & 0.029 & 0.094 & 6600 & ...  \\
{\objd} &   15.990 &  0.004 & 0.43   & 0.006 & 14.707 &  0.077 & 0.024 & 0.028 & 6302 & 6097 \\
{\obje} &   15.480 &  0.004 & 0.42   & 0.006 & 14.165 &  0.053 & 0.014 &\nodata& 6184 & 6097 \\
{\objf} &   15.810 &  0.004 & 0.33   & 0.006 & 14.776 &  0.108 & 0.035 & 0.091 & 6700 & 6600 \\
{\objb} &   16.009 &  0.004 & 0.390  & 0.006 & 14.880 &  0.120 & 0.088 & 0.126 & 6800 & 6600 \\ %
\enddata
\tablenotetext{a}{$V$ and $B-V$ for SDSS/SEGUE objects are already extinction
and reddening corrected}
\tablenotetext{b}{$E(B-V)$ from the dust map of \citet{schlegel98}}
\tablenotetext{c}{$E(B-V)$ from the \ion{Na}{1} D1 line}
\end{deluxetable}


\begin{deluxetable}{lccccc}
\tablewidth{0pt}
\tablecaption{RADIAL VELOCITY VARIATION FOUND FOR {\objf}  \label{tab:s0353}}
\tablehead{
UT  & HJD & exposure & Shift & Width &  $V_{\rm helio}$ \\ 
    &     & (minutes) & ({\kms}) & (m{\AA}) & ({\kms}) 
}
\startdata
10 Feb.2007, 14:19 & 2454142.097 & 40 & 35.4 & 27.5 &  40.7 \\
10 Feb.2007, 15:00 & 2454142.125 & 40 & 16.9 & 35.5 &  22.2 \\
10 Feb.2007, 15:41 & 2454142.153 & 37 &$-$2.3 & 23.3 &   3.0 \\
11 Feb.2007, 15:34 & 2454143.148 & 20 & 15.6 & 16.9 &  20.8 \\
11 Feb.2007, 15:49 & 2454143.159 & 20 & 16.5 & 20.0 &  21.7 \\
\enddata
\end{deluxetable}

\clearpage
\pagestyle{empty}
\begin{deluxetable}{lccccccccccc}
\tablewidth{0pt}
\tabletypesize{\scriptsize}
\rotate
\tablecaption{EQUIVALENT WIDTHS \label{tab:ew}}
\tablehead{Species &   Wavelength & L.E.P.&  $\log gf$ & \multicolumn{8}{c}{Equivalent width (m{\AA})} \\
        &   ({\AA})    & (eV) &         & {\LP} & {\CS} & {\obja}  & {\objc}  & {\objd}  & {\obje}  & {\objf}  & {\objb} 
}
\startdata
     Na I &   5889.95&    0.00&    0.10&   164.2&    98.6&   116.1&    75.3&    ... &   147.3&  313.8 &    69.5 \\
     Na I &   5895.92&    0.00&   -0.20&   135.5&    77.0&    96.4&    55.5&    ... &   127.3&  277.6 &    ...  \\
     Mg I &   4057.50&    4.35&   -0.89&    13.0&    ... &    ... &    ... &    ... &    ... &    ... &    ...  \\
     Mg I &   4571.10&    0.00&   -5.69&     1.9&    ... &    ... &    ... &    ... &    ... &    ... &    ...  \\
     Mg I &   5172.69&    2.71&   -0.38&   129.6&   117.6&    99.8&    84.9&    71.8&   120.2&   184.2&   124.9 \\
     Mg I &   5183.60&    2.72&   -0.16&   147.5&   135.2&   122.0&   101.6&    77.3&   141.4&   196.5&   147.3 \\
     Mg I &   5528.40&    4.35&   -0.49&    33.0&    28.8&    22.7&    11.2&    22.0&    28.7&    ... &    40.7 \\
     Ca I &   4226.73&    0.00&    0.24&    ... &    ... &    ... &    ... &    ... &    ... &   122.1&    ...  \\
     Ca I &   4435.69&    1.89&   -0.52&    13.8&    17.9&     7.1&    ... &    ... &    ... &    ... &    27.7 \\
     Ca I &   4454.78&    1.90&    0.26&    35.8&    41.0&    38.0&    ... &    ... &    41.2&    ... &    49.3 \\
     Ca I &   4455.89&    1.90&   -0.53&    10.3&    12.4&     8.7&    ... &    ... &    12.9&    ... &    13.0 \\
     Ca I &   5265.56&    2.52&   -0.11&     6.3&    ... &     6.8&    ... &    ... &     7.3&    ... &    ...  \\
     Ca I &   5588.76&    2.53&    0.36&    12.5&    ... &    13.0&    ... &    ... &    12.8&    ... &    ...  \\
     Ca I &   5594.47&    2.52&    0.10&    13.5&    17.4&    10.6&    ... &    ... &    ... &    ... &    13.0 \\
     Ca I &   5598.49&    2.52&   -0.09&     6.3&    ... &     7.5&    ... &    ... &    ... &    ... &    19.7 \\
     Ca I &   5857.45&    2.93&    0.24&     5.9&    ... &    ... &    ... &    ... &    ... &    ... &    18.3 \\
     Ca I &   6102.72&    1.88&   -0.77&     4.3&    ... &    ... &    ... &    ... &    ... &    ... &    15.7 \\
     Ca I &   6122.22&    1.89&   -0.32&    12.7&    16.5&    ... &    ... &    ... &    ... &    ... &    21.1 \\
     Ca I &   6162.17&    1.90&   -0.09&    19.2&    22.3&    21.5&    ... &    ... &    ... &    ... &    ...  \\
     Ca I &   6439.07&    2.53&    0.39&    16.8&    19.1&    ... &    ... &    ... &    17.7&    ... &    26.2 \\
     Ca I &   6462.57&    2.52&    0.26&    13.0&    13.1&    13.1&    ... &    ... &    16.5&    ... &    ...  \\
     Ti I &   4981.73&    0.85&    0.56&     9.9&     9.1&    ... &    ... &    ... &    ... &    ... &    ...  \\
     Ti I &   4991.07&    0.84&    0.44&     6.8&    ... &    ... &    ... &    ... &    ... &    ... &    ...  \\
     Ti I &   4999.50&    0.83&    0.31&     8.2&    11.9&    ... &    ... &    ... &    ... &    ... &    ...  \\
     Ti I &   5007.21&    0.82&    0.17&     8.6&    ... &    ... &    ... &    ... &    ... &    ... &    14.4 \\
     Ti I &   5064.65&    0.05&   -0.94&     4.6&    ... &    ... &    ... &    ... &    ... &    ... &    ...  \\
     Cr I &   4652.16&    1.00&   -1.03&     4.1&    ... &    ... &    ... &    ... &    ... &    ... &    ...  \\
     Cr I &   5206.04&    0.94&    0.02&    21.0&    ... &    14.6&    ... &    ... &    13.4&    ... &    26.2 \\
     Cr I &   5208.44&    0.94&    0.16&    23.5&    31.9&    22.5&    ... &    ... &    25.9&    ... &    37.4 \\
     Fe I &   4063.59&    1.56&    0.06&    88.2\tablenotemark{a} &    ... &    ... &    ... &    ... &    ... &    ... &    ...  \\
     Fe I &   4071.74&    1.61&   -0.02&    82.1&    90.4&    75.8&    ... &    56.5&    74.8&    ... &    74.5 \\
     Fe I &   4107.49&    2.83&   -0.88&    10.6&    20.7&    ... &    ... &    ... &    ... &    ... &    ...  \\
     Fe I &   4143.41&    3.05&   -0.20&    17.9&    19.3&    19.2&    ... &    ... &    27.0&    ... &    23.8 \\
     Fe I &   4143.87&    1.56&   -0.51&    61.5&    68.6&    59.2&    26.9&    ... &    62.1&    ... &    59.6 \\
     Fe I &   4202.03&    1.49&   -0.71&    ... &    ... &    ... &    ... &    ... &    ... &    43.3&    ...  \\
     Fe I &   4271.76&    1.49&   -0.16&    ... &    ... &    ... &    ... &    ... &    ... &    38.1&    ...  \\
     Fe I &   4307.90&    1.56&   -0.07&    ... &    ... &    ... &    ... &    ... &    ... &    62.1&    ...  \\
     Fe I &   4325.76&    1.61&    0.01&    ... &    ... &    ... &    ... &    ... &    ... &    65.0&    ...  \\
     Fe I &   4383.54&    1.49&    0.20&    98.9\tablenotemark{a} &   102.5&    87.1&    69.4&    ... &   102.2&    91.0&    99.6 \\
     Fe I &   4404.75&    1.56&   -0.14&    77.6&    84.7&    65.4&    56.4&    41.5&    52.6&    66.7&    76.9 \\
     Fe I &   4415.12&    1.61&   -0.62&    67.5&    65.4&    59.5&    22.7&    21.0&    85.3&    37.2&    65.9 \\
     Fe I &   4427.31&    0.05&   -2.92&    36.0&    19.7&    29.3&    ... &    ... &    68.0&    ... &    39.7 \\
     Fe I &   4442.34&    2.20&   -1.25&    ... &    19.4&    22.3&    ... &    ... &    ... &    ... &    19.5 \\
     Fe I &   4447.72&    2.22&   -1.34&    ... &     7.1&    ... &    ... &    ... &    ... &    ... &    ...  \\
     Fe I &   4459.12&    2.18&   -1.28&    14.2&    15.6&    11.8&    ... &    ... &    14.9&    ... &    22.1 \\
     Fe I &   4461.65&    0.09&   -3.21&    13.2&    ... &    ... &    ... &    ... &    12.5&    ... &    22.1 \\
     Fe I &   4466.55&    2.83&   -0.60&    14.5&    22.9&    17.4&    17.8\tablenotemark{a} &    ... &    12.6&    ... &    ...  \\
     Fe I &   4476.02&    2.85&   -0.82&    13.3&    ... &     7.5&    ... &    ... &    ... &    ... &    11.6 \\
     Fe I &   4494.56&    2.20&   -1.14&    15.3&     9.0&    11.9&    ... &    ... &    15.2&    ... &    ...  \\
     Fe I &   4528.61&    2.18&   -0.82&    29.6&    42.7&    22.6&    ... &    ... &    29.8&    ... &    28.7 \\
     Fe I &   4531.15&    1.49&   -2.15&    10.7&    ... &    ... &    ... &    ... &    ... &    ... &    ...  \\
     Fe I &   4592.65&    1.56&   -2.45&     4.9&    ... &    ... &    ... &    ... &    ... &    ... &    ...  \\
     Fe I &   4602.94&    1.49&   -2.21&     6.8&    ... &    ... &    ... &    ... &    ... &    ... &    11.5 \\
     Fe I &   4871.32&    2.87&   -0.36&    ... &    17.2&    ... &    ... &    ... &    ... &    ... &    ...  \\
     Fe I &   4872.14&    2.88&   -0.57&    13.3&    12.9&     9.6&    ... &    ... &    ... &    ... &    16.1 \\
     Fe I &   4890.75&    2.88&   -0.39&    22.2&    28.9&    11.1&    ... &    ... &    19.5&    ... &    28.4 \\
     Fe I &   4891.49&    2.85&   -0.11&    31.8&    34.5&    21.9&    ... &    ... &    36.5&    ... &    31.5 \\
     Fe I &   4903.31&    2.88&   -0.93&     5.2&     9.4&    ... &    ... &    ... &    ... &    ... &    17.4 \\
     Fe I &   4918.99&    2.87&   -0.34&    19.7&    24.4&    ... &    ... &    ... &    21.1&    ... &    28.8 \\
     Fe I &   4920.50&    2.83&    0.07&    40.3&    54.7&    31.8&    ... &    ... &    40.2&    ... &    46.5 \\
     Fe I &   4957.30&    2.85&   -0.41&    22.7&    ... &    ... &    ... &    ... &    21.8&    ... &    ...  \\
     Fe I &   4957.60&    2.81&    0.23&    49.7&    61.9&    42.2&    ... &    23.8&    42.2&    ... &    ...  \\
     Fe I &   4966.09&    3.33&   -0.87&     5.1&     6.2&    ... &    ... &    ... &    ... &    ... &    ...  \\
     Fe I &   4994.13&    0.92&   -2.96&     4.2&    ... &    ... &    ... &    ... &    ... &    ... &    ...  \\
     Fe I &   5006.12&    2.83&   -0.61&    13.5&    12.9&    ... &    ... &    ... &    14.2&    ... &    ...  \\
     Fe I &   5012.07&    0.86&   -2.64&    10.7&    ... &     8.6&    ... &    ... &    ... &    ... &    19.9 \\
     Fe I &   5041.76&    1.49&   -2.20&    ... &    28.8\tablenotemark{a} &    11.4&    ... &    ... &    ... &    ... &    ...  \\
     Fe I &   5049.82&    2.28&   -1.34&    10.0&     9.9&    ... &    ... &    ... &    14.8&    ... &    ...  \\
     Fe I &   5051.63&    0.92&   -2.80&     6.5&    ... &    ... &    ... &    ... &    15.2\tablenotemark{a} &    ... &    10.0 \\
     Fe I &   5151.91&    1.01&   -3.32&     7.7\tablenotemark{a} &    ... &    ... &    ... &    ... &    ... &    ... &    ...  \\
     Fe I &   5171.60&    1.49&   -1.79&    16.8&    14.3&    ... &    ... &    ... &    13.8&    ... &    23.2 \\
     Fe I &   5191.46&    3.04&   -0.55&    10.7&    26.5&     9.0&    ... &    ... &    13.1&    ... &    17.2 \\
     Fe I &   5192.34&    3.00&   -0.42&    14.4\tablenotemark{a} &    15.7&    11.0&    ... &    ... &    16.7&    ... &    20.8 \\
     Fe I &   5194.94&    1.56&   -2.09&     9.5&    ... &    14.6&    ... &    ... &     8.9&    ... &    ...  \\
     Fe I &   5198.71&    2.22&   -2.13&     2.2&    ... &    ... &    ... &    ... &    ... &    ... &    ...  \\
     Fe I &   5202.34&    2.18&   -1.84&     4.4&    ... &    ... &    ... &    ... &    ... &    ... &    ...  \\
     Fe I &   5216.27&    1.61&   -2.15&     7.5&    ... &    ... &    ... &    ... &    ... &    ... &    ...  \\
     Fe I &   5232.94&    2.94&   -0.06&    28.7&    32.6&    19.5&    ... &    ... &    29.6&    ... &    39.2 \\
     Fe I &   5266.56&    3.00&   -0.39&    14.9&    20.5&    14.3&    ... &    ... &    15.2&    ... &    16.9 \\
     Fe I &   5269.54&    0.86&   -1.32&    67.4&    64.6&    57.4&    17.7&    42.4&    58.0&   39.4 &    74.5 \\
     Fe I &   5270.36&    1.61&   -1.34&    36.0&    40.9&    29.8&    ... &    ... &    37.3&    ... &    57.5\tablenotemark{a}  \\
     Fe I &   5281.79&    3.04&   -0.83&     4.7&    ... &    ... &    ... &    ... &    ... &    ... &    ...  \\
     Fe I &   5324.18&    3.21&   -0.10&    16.4&    23.8&    14.4&    ... &    ... &    17.7&    ... &    23.8 \\
     Fe I &   5328.04&    0.92&   -1.47&    57.4&    57.9&    45.9&    13.4&    ... &    50.4&   40.6 &    67.7 \\
     Fe I &   5328.53&    1.56&   -1.85&    14.4&    13.1&    11.5&    ... &    ... &    10.2&    ... &    19.7 \\
     Fe I &   5339.93&    3.27&   -0.65&     7.9&    ... &    ... &    ... &    ... &    ... &    ... &    ...  \\
     Fe I &   5455.61&    1.01&   -2.10&    27.5&    30.7&    22.1&    ... &    ... &    ... &    ... &    34.9 \\
     Fe I &   5497.52&    1.01&   -2.85&     4.9&    ... &    ... &    ... &    ... &    ... &    ... &    ...  \\
     Fe I &   5506.78&    0.99&   -2.80&     6.5&    ... &    ... &    ... &    ... &    ... &    ... &    ...  \\
     Fe I &   5569.62&    3.42&   -0.54&     6.2&    10.7&    ... &    ... &    ... &    11.1&    ... &    ...  \\
     Fe I &   5572.84&    3.40&   -0.28&     8.7&    13.9&    ... &    ... &    ... &    13.0&    ... &    15.8 \\
     Fe I &   5586.75&    3.37&   -0.10&    12.6&    17.2&     8.5&    ... &    ... &    14.4&    ... &    17.5 \\
     Fe I &   5615.64&    3.33&    0.05&    18.2&    23.5&    14.2&    ... &    ... &    20.9&    ... &    24.1 \\
     Fe I &   6136.61&    2.45&   -1.40&     6.9&    ... &    ... &    ... &    ... &    ... &    ... &    ...  \\
     Fe I &   6137.69&    2.59&   -1.40&     6.9&    ... &    ... &    ... &    ... &    ... &    ... &    ...  \\
     Fe I &   6191.56&    2.43&   -1.42&     6.4&    ... &    ... &    ... &    ... &    ... &    ... &    ...  \\
     Fe I &   6230.72&    2.56&   -1.28&     6.2&    ... &    ... &    ... &    ... &     9.4&    ... &    ...  \\
     Fe I &   6393.60&    2.43&   -1.43&     5.8&    ... &    ... &    ... &    ... &    ... &    ... &    ...  \\
     Fe I &   6677.99&    2.69&   -1.42&     4.8&    ... &    ... &    ... &    ... &    ... &    ... &    ...  \\
     Co I &   4121.32&    0.92&   -0.32&    10.6&    ... &     7.7&    ... &    ... &    ... &    ... &    ...  \\
     Ni I &   5476.91&    1.83&   -0.89&    11.9&    21.0&     5.7&    ... &    ... &    11.9&    ... &    16.5 \\
     Zn I &   4722.15&    0.00&   -0.37&     3.0&    ... &    ... &    ... &    ... &    ... &    ... &    ...  \\
     Sc II&   4320.75&    0.61&   -0.25&    11.5&    ... &    ... &    ... &    ... &    19.4&    ... &    ...  \\
     Ti II&   4395.00&    1.08&   -0.51&    ... &    ... &    ... &    ... &    ... &    ... &    42.3&    ...  \\
     Ti II&   4443.77&    1.08&   -0.70&    35.8&    45.6&    33.3&    18.1&    ... &    42.2&    ... &    46.1 \\
     Ti II&   4444.54&    1.12&   -2.21&    ... &    ... &     5.1&    ... &    ... &    ... &    ... &    ...  \\
     Ti II&   4450.50&    1.08&   -1.51&     6.9&    12.2&    ... &    ... &    ... &    ... &    ... &    14.3 \\
     Ti II&   4464.46&    1.16&   -2.08&     5.4&    ... &    ... &    ... &    ... &     6.6&    ... &    ...  \\
     Ti II&   4468.52&    1.13&   -0.60&    39.4&    52.7&    37.6&    29.3&    ... &    38.1&    ... &    57.5 \\
     Ti II&   4501.27&    1.12&   -0.76&    29.8&    44.4&    29.2&    20.1&    ... &    29.8&    ... &    47.0 \\
     Ti II&   4533.97&    1.24&   -0.77&    37.1&    47.6&    34.6&    ... &    ... &    37.7&    33.2&    57.0 \\
     Ti II&   4563.77&    1.22&   -0.96&    24.5&    29.7&    29.8&    13.9&    ... &    24.3&    31.2&    27.6 \\
     Ti II&   4571.96&    1.57&   -0.53&    29.6&    ... &    33.9&    ... &    ... &    41.8&    33.0&    ...  \\
     Ti II&   4589.92&    1.24&   -1.79&    ... &    ... &    ... &    ... &    ... &    ... &    ... &    11.1 \\
     Ti II&   4805.09&    2.06&   -1.10&    ... &     9.6&    ... &    ... &    ... &     8.7&    ... &    ...  \\
     Ti II&   5226.53&    1.57&   -1.30&     6.5&    ... &     9.5&    ... &    ... &     8.3&    ... &    16.0 \\
     Fe II&   4491.40&    2.86&   -2.70&     3.9&    ... &    ... &    ... &    ... &    ... &    ... &    10.6 \\
     Fe II&   4508.28&    2.86&   -2.58&     7.4&    13.4&    ... &    ... &    ... &    ... &    ... &    12.8 \\
     Fe II&   4515.34&    2.84&   -2.48&    ... &     9.9&    ... &    ... &    ... &    ... &    ... &    ...  \\
     Fe II&   4520.23&    2.81&   -2.60&     4.5&     9.1&    ... &    ... &    ... &    ... &    ... &    ...  \\
     Fe II&   4522.63&    2.84&   -2.03&    10.0&    ... &    11.9&    ... &    ... &    17.5&    ... &    17.9 \\
     Fe II&   4555.89&    2.83&   -2.29&    ... &    20.3&    ... &    ... &    ... &    ... &    ... &    17.1 \\
     Fe II&   4583.83&    2.81&   -2.02&    17.1&    42.7&    ... &    ... &    ... &    20.1&    22.1&    26.8 \\
     Fe II&   4923.93&    2.89&   -1.32&    34.2&    63.4&    32.6&    15.7&    ... &    ... &    23.9&    46.8 \\
     Fe II&   5018.45&    2.89&   -1.22&    41.4&    68.5&    43.1&    21.7&    ... &    45.4&    47.3&    57.2 \\
     Fe II&   5197.56&    3.23&   -2.10&     3.5&    14.1&     8.1&    ... &    ... &    ... &    ... &    ...  \\
     Fe II&   5234.62&    3.22&   -2.27&     4.8&    12.9&    ... &    ... &    ... &    ... &    ... &    ...  \\
     Fe II&   5276.00&    3.20&   -1.94&     7.4&    18.0&    10.9&    ... &    ... &    ... &    ... &    17.7 \\
     Fe II&   5316.62&    3.15&   -1.85&     9.5&    29.3&    ... &    ... &    ... &    10.6&    ... &    17.0 \\
     Sr II&   4077.71&    0.00&    0.15&    74.1&   129.9&    60.4&   102.7&    ... &    91.5&   287.1&   104.5 \\
     Sr II&   4215.52&    0.00&   -0.18&    63.5&   111.0&    43.9&    92.6&    43.7&    82.8&   160.9&    99.0 \\
      Y II&   4204.69&    0.00&   -1.76&    61.5&    ... &    55.5&    25.2&    ... &   101.2&    ... &    ...  \\
      Y II&   4883.68&    1.08&    0.07&    ... &    15.1&    ... &    ... &    ... &    ... &    ... &    12.7 \\
      Y II&   4900.12&    1.03&   -0.09&    ... &    47.1&    ... &    ... &    ... &    ... &    35.6&    ...  \\
     Zr II&   4048.67&    0.80&   -0.48&    ... &    14.8&    ... &    ... &    ... &    ... &    ... &    ...  \\
     Zr II&   4149.20&    0.80&   -0.03&    ... &    31.9&    ... &    20.0&    ... &    32.6&    ... &    24.2 \\
     Zr II&   4150.97&    0.80&   -1.08&     4.2&    ... &    ... &    ... &    ... &    ... &    ... &    ...  \\
     Ba II&   4554.03&    0.00&    0.16&   152.3&    ... &    ... &   140.8&    51.2&    88.9&   282.2&   108.1 \\
     Ba II&   4934.09&    0.00&   -0.16&   147.0\tablenotemark{a} &   169.5&    34.1&   127.9&    ... &   126.9&   248.2&   102.8 \\
     Ba II&   5853.70&    0.60&   -1.01&    48.1&    77.8&     9.0&    38.6&    ... &    37.8&    77.6&    30.2 \\
     Ba II&   6141.70&    0.70&   -0.07&    92.4&   123.4&    12.7&    85.3&    ... &    78.4&   192.0&    74.4 \\
     Ba II&   6496.91&    0.60&   -0.38&    81.8&   115.8&     8.6&    88.6&    ... &    68.3&   150.3&    63.5 \\
     La II&   4086.71&    0.00&   -0.07&    23.1&    41.5&    ... &    16.5&    ... &    ... &    ... &     8.0 \\
     La II&   4123.22&    0.32&    0.13&    21.4&    43.2&    ... &    16.7&    ... &    ... &    ... &    12.7 \\
\hline
     Na I\tablenotemark{b} &   5889.95&    0.00&    0.10&    27.6&    90.6&    38.9&   259.0&   252.0&   100.0&    ... &   319.0 \\
\enddata
\tablenotetext{a}{Equivalent width measured, but not used in the analysis.}
\tablenotetext{b}{Equivalent width of interstellar absorption.}
\end{deluxetable}

\clearpage
\thispagestyle{plaintop}
\begin{deluxetable}{llll}
\tablewidth{0pt}
\tabletypesize{\footnotesize}
\tablecaption{\label{tab:rv_sdss} RADIAL VELOCITIES FROM SDSS SPECTRA}
\tablehead{
Star    & JD &   $V_{\rm helio}$ ({\kms}) & Remarks 
}
\startdata
{\obja} & 2,452,146 & $-150.2 \pm 2.4$ & SDSS \\
        & 2,452,162 & $-143.9 \pm 1.9$ & SDSS \\
        & 2,453,993 & $-146.18 \pm 0.18$ & this work \\
{\objc} & 2,453,712 & $-267.1 \pm  2.5$ &  SDSS \\
        & 2,453,713 & $-273.8 \pm  2.9$ & SDSS \\
        & 2,453,993 & $-272.24 \pm 0.28$ & this work \\
{\objd} & 2,452,709 & $45.8 \pm 3.5$ & SDSS \\ 
        & 2,454,142 & $1.7 \pm 2.5$ & this work \\
{\obje} & 2,452,708 & $-365.5 \pm 1.8 $ & SDSS \\ 
        & 2,452,636 & $-369.5 \pm 2.2 $ & SDSS \\ 
        & 2,454,142 & $-366.16 \pm 0.23$ & this work \\
{\objf} & 2,451,703 & $37.2 \pm 3.2$ & SDSS \\
{\objb} & 2,452,466 & $-419.2  \pm 2.3$ & SDSS \\
        & 2,452,524 & $-420.4  \pm 1.9$ &SDSS \\
        & 2,452,932 & $-404.3  \pm 2.3$ &SDSS \\
        & 2,453,993 & $-417.92 \pm 0.20$ & this work\\
\enddata
\end{deluxetable}

\begin{deluxetable}{lcccccccc}
 \tablewidth{0pt}
\tablecaption{ATMOSPHERIC PARAMETERS \label{tab:param}}
\tablehead{
Star    & {\teff} & $\sigma$({\teff}) & {\logg} & $\sigma$({\logg}) & [Fe/H] & $\sigma$([Fe/H]) & {\vt} & $\sigma$({\vt}) \\
        & (K)     & (K)               & (dex) & (dex) & (dex) & (dex)  & {\kms} & {\kms}  
}
\startdata
{\LP}    & 6200 & 150 & 4.3 & 0.3 & $-2.5$ & 0.3 & 1.4 & 0.3 \\
{\CS}    & 6800 & 150 & 4.1 & 0.3 & $-2.1$ & 0.3 & 2.1 & 0.3 \\
{\obja}  & 6500 & 200 & 4.5 & 0.3 & $-2.5$ & 0.3 & 1.5 & 0.3 \\
{\objc}  & 6600 & 200 & 4.1 & 0.3 & $-3.2$ & 0.3 & 2.0 & 0.5 \\
{\objd}  & 6300 & 150 & 4.0 & 0.5 & $-3.2$ & 0.3 & 1.5 & 0.5 \\
{\obje}  & 6200 & 150 & 4.0 & 0.3 & $-2.6$ & 0.3 & 1.4 & 0.3 \\
{\objf}  & 6700 & 200 & 4.2 & 0.3 & $-2.5$ & 0.3 & 1.5 & 0.5 \\
{\objb}  & 6600 & 200 & 4.5 & 0.3 & $-2.1$ & 0.3 & 1.3 & 0.3 \\
\enddata
\end{deluxetable}

\clearpage

\begin{deluxetable}{@{}l@{\extracolsep{\fill}}c@{\extracolsep{\fill}}c@{\extracolsep{\fill}}c@{\extracolsep{\fill}}c@{\extracolsep{\fill}}c@{\extracolsep{\fill}}c@{\extracolsep{\fill}}c@{\extracolsep{\fill}}c@{\extracolsep{\fill}}c@{\extracolsep{\fill}}c@{\extracolsep{\fill}}c@{\extracolsep{\fill}}c@{\extracolsep{\fill}}c@{\extracolsep{\fill}}c@{\extracolsep{\fill}}c@{\extracolsep{\fill}}c@{\extracolsep{\fill}}c@{\extracolsep{\fill}}c@{\extracolsep{\fill}}c@{\extracolsep{\fill}}c@{\extracolsep{\fill}}c}
\tabletypesize{\scriptsize}
\rotate
\tablewidth{24cm}
\tablecaption{\label{tab:abund} CHEMICAL ABUNDANCES RESULTS}
\tablehead{
     & FeI & FeII & Li & C & Na I & MgI & CaI & ScII & TiI & TiII & CrI & CoI & NiI & ZnI & SrII & YII & ZrII & BaII & LaII & Pb I 
}
\startdata
Sun & &&&&&&&&&&&&&&&&&&&& \\
\hline
 log(A) & 7.45 & 7.45 & 1.05 & 8.39 & 6.17 & 7.53 & 6.31 & 3.05 & 4.90 & 4.90 & 5.64 & 4.92 & 6.23 & 4.60 & 2.92 & 2.21 & 2.59 & 2.17 & 1.13 & 2.00 \\

\hline
{\LP} & &&&&&&&&&&&&&&&&&&&& \\
\hline
$\log \epsilon$(X) & 4.92 & 4.92 & 2.3 & 8.00 & 4.76 & 5.42 & 4.00 & 0.72 & 2.92 & 2.76 & 3.04 & 2.60 & 3.77 & 2.07 & 0.47 & \nodata & \nodata & 1.72 & 0.52 & 2.0 \\
$[$X/Fe] & -2.53 & -2.53 & \nodata & 2.14 & 1.12 & 0.42 & 0.22 & 0.20 & 0.55 & 0.39 & -0.07 & 0.21 & 0.07 & 0.00 & 0.08 & \nodata & \nodata & 2.08 & 1.92 & 2.53 \\
$ N$   &   59 &   11 &  1  &     & 2    & 5    & 13   & 1    & 5    & 9    & 3    & 1    & 1    & 1    & 2    & \nodata   & \nodata   & 4     & 2     & 1   \\
$\sigma$ & 0.12 & 0.12 & 0.2 & 0.25 & 0.25 & 0.13 & 0.09 & 0.18 & 0.14 & 0.12 & 0.14 & 0.19 & 0.18 & 0.13 & 0.22 & \nodata  &  \nodata & 0.18 & 0.16 & \\

\hline
{\CS} & &&&&&&&&&&&&&&&&&&&& \\
\hline
$\log \epsilon$(X) & 5.39 & 5.38 & $<2.3$ & 8.4 & 4.25 & 5.69 & 4.45 & \nodata & 3.39 & 3.08 & 3.53 &  & 4.52 & \nodata & 1.63 & 1.50 & 1.80 & 2.5 & 1.16 & 3.1 \\
$[$X/Fe] & -2.06 & -2.07  && 2.07 & 0.14 & 0.22 & 0.19 & \nodata & 0.54 & 0.24 & -0.05 & \nodata & 0.35 & \nodata & 0.77 & 1.34 & 1.26 & 2.39 & 2.09 & 3.16 \\
$N$ & 40 & 11 & &  & 2 & 4 & 8 & & 2 & 7 & 1 & \nodata & 1 & \nodata  & 2 & 2 & 2 & 4 & 2 & 1\\
$\sigma$ & 0.11 & 0.12 & & 0.26 & 0.19 & 0.16 & 0.10 & \nodata & 0.11 & 0.14 & 0.11 & \nodata & 0.11 & \nodata & 0.27 & 0.20 & 0.20 & 0.22 & 0.20 & \\

\hline
{\obja} & &&&&&&&&&&&&&&&&&&&& \\
\hline
$\log \epsilon$(X) & 5.04 & 5.03 & $<2.0$ &8.30 & 4.54 & 5.37 & 4.16 & \nodata & \nodata & 3.00 & 3.05 & 2.73 & 3.65 & \nodata & 0.24 & \nodata & \nodata & 0.05 & \nodata & $<2.3$ \\
$[$X/Fe] & -2.41 & -2.42 & & 2.32 & 0.78 & 0.25 & 0.26 & \nodata & \nodata & 0.51 & -0.18 & 0.22 & -0.17 & \nodata & -0.27 & \nodata & \nodata & 0.29 & \nodata & \\
$N$ & 33 & 5 & &  & 2 & 3 & 9 & \nodata & \nodata & 8 & 2 & 1 & 1  & \nodata & 2 & \nodata & \nodata & 4 & \nodata &  \\
$\sigma$ & 0.17 & 0.15 & & 0.32 & 0.32 & 0.19 & 0.13 & \nodata & \nodata & 0.15 & 0.16 & 0.18 & 0.15 & \nodata & 0.29 & \nodata & \nodata & 0.24 & \nodata &  \\

\hline
{\objc} & &&&&&&&&&&&&&&&&&&&& \\
\hline
$\log \epsilon$(X) & 4.34 & 4.34 & $<2.2$ & 8.2 & 3.75 & 5.03 & \nodata &  & \nodata & 2.45 & \nodata &  & \nodata &   & 1.15 & \nodata & 1.41 & 1.81 & 0.48 & 2.3\\
$[$X/Fe] & -3.11 & -3.11 & & 2.92 & 0.69 & 0.61 & \nodata & \nodata & \nodata & 0.66 & \nodata & \nodata & \nodata & \nodata & 1.35 & \nodata & 1.93 & 2.75 & 2.46 & 3.41\\
$N$ & 6 & 2 & &  & 2 & 4 & \nodata & \nodata & \nodata & 4 & \nodata & \nodata & \nodata & \nodata  & 2 & \nodata & 1 & 5 & 2 & 1\\
$\sigma$ & 0.16 & 0.14 & & 0.32 & 0.20 & 0.18 & \nodata & \nodata & \nodata & 0.14 & \nodata & \nodata & \nodata & \nodata & 0.37 & \nodata & 0.19 & 0.30 & 0.16 &  \\

\hline
{\objd} & &&&&&&&&&&&&&&&&&&&& \\
\hline
$\log \epsilon$(X) & 4.29 & \nodata & $<2.3$ &$<7.6$ & \nodata & 4.8 & \nodata & \nodata & \nodata & \nodata & \nodata &  & \nodata &   & -0.1 & \nodata & \nodata & -0.22 &  & \\
$[$X/Fe]           & -3.16 & \nodata & & $<2.2$ & \nodata & 0.43 & \nodata & \nodata & \nodata & \nodata & \nodata & \nodata & \nodata & \nodata & 0.14 & \nodata & \nodata & 0.77 & \nodata & \\
$N$                & 5 & \nodata &  & & \nodata & 3 & \nodata & \nodata & \nodata & \nodata & \nodata & \nodata & \nodata & \nodata & 1 & \nodata & \nodata & 1 & \nodata & \nodata\\
$\sigma$           & 0.20 & \nodata & & \nodata & \nodata & 0.23 & \nodata & \nodata & \nodata & \nodata & \nodata & \nodata & \nodata & \nodata & 0.40 & \nodata & \nodata & 0.35 & \nodata & \nodata\\

\hline
{\obje} & &&&&&&&&&&&&&&&&&&&& \\
\hline
$\log \epsilon$(X) & 4.94 & 4.91 & $<2.0$ & 8.6 & 4.97 & 5.55 & 4.08 & 0.89 & \nodata & 2.79 & 2.84 &  & 3.78 & \nodata  & 1.02 & \nodata & \nodata & 1.48 & \nodata & 2.5\\
$[$X/Fe] & -2.51 & -2.55 & & 2.72 & 1.31 & 0.52 & 0.28 & 0.35 & \nodata & 0.40 & -0.29 & \nodata & 0.05 & \nodata & 0.60 & \nodata & \nodata & 1.81 & \nodata & 3.01\\
$N$ & 33 & 4 & &  & 2 & 4 & 6 & 1 & \nodata & 9 & 2 &  & 1 & \nodata & 2 & \nodata & \nodata & 5 & \nodata & 1 \\
$\sigma$ & 0.16 & 0.13 & & 0.32 & 0.29 & 0.15 & 0.12 & 0.14 & \nodata & 0.14 & 0.16 & \nodata & 0.15 & \nodata & 0.25 & \nodata & \nodata & 0.22 & \nodata & \\

\hline
{\objf} & &&&&&&&&&&&&&&&&&&&& \\
\hline
$\log \epsilon$(X) & 4.93 & 4.96 & $<2.5$ & 8.0: & 6.36 & 6.14 & 4.58 & \nodata & \nodata & 2.98 & \nodata &  & \nodata & \nodata  & 2.65 & \nodata & \nodata & 3.05 &  & $<3.2$\\
$[$X/Fe] & -2.52 & -2.49 & & +2.1: & 2.71 & 1.13 & 0.79 & \nodata & \nodata & 0.60 & \nodata & \nodata & \nodata & \nodata & 2.25 & \nodata & \nodata & 3.40 & \nodata & \\
$N$ & 9 & 3 & &  & 2 & 2 & 1 & \nodata & \nodata & 4 & \nodata & \nodata & \nodata & \nodata  & 2 & \nodata & \nodata & 5 & \nodata & \\
$\sigma$ & 0.16 & 0.15 & & 0.32 & 0.22 & 0.21 & 0.21 & \nodata & \nodata & 0.15 & \nodata & \nodata & \nodata & \nodata & 0.38 & \nodata & \nodata & 0.31 & \nodata & \\

\hline
{\objb} & &&&&&&&&&&&&&&&&&&&& \\
\hline
$\log \epsilon$(X) & 5.40 & 5.40 & $<2.3$ &   & 4.45 & 5.75 & 4.59 & \nodata & 3.66 & 3.29 & 3.49 & \nodata & 4.26 & \nodata  & 1.56 & 0.95 & 1.5 & 1.62 & \nodata & \nodata\\
$[$X/Fe]           & -2.05 & -2.05 & & 2.00 & 0.33 & 0.27 & 0.32 & \nodata & 0.80 & 0.44 & -0.10 & \nodata & 0.08 & \nodata & 0.68 & 0.79 & 0.96 & 1.50 & \nodata & \nodata\\
$N$ & 32 & 9 & &  & 1 & 4 & 9 & \nodata & 1 & 8 & 2 & \nodata & 1 & \nodata & 2 & 1 & 1 & 5 & \nodata & \nodata\\
$\sigma$ & 0.15 & 0.13 & & 0.32 & 0.28 & 0.18 & 0.13 & \nodata & 0.15 & 0.15 & 0.15 & \nodata & 0.15 & \nodata & 0.29 & \nodata & \nodata & 0.24 & \nodata & \nodata\\

\enddata
\end{deluxetable}

\clearpage
\thispagestyle{plaintop}
\begin{deluxetable}{lcccccccccc}
\tabletypesize{\small}
 \tablewidth{0pt}
\tablecaption{ABUNDANCE CHANGES BY CHANGING ATMOSPHERIC PARAMETERS \label{tab:err}}
\tablehead{
    & \multicolumn{5}{c}{\LP}  & & \multicolumn{4}{c}{\CS} \\
  \cline{2-6} \cline{8-11}
    & $\sigma${\teff} &  $\sigma${\logg} & $\sigma$[Fe/H] & $\sigma${\vt}  & $\Delta_{\rm ATLAS}$ &  & $\sigma${\teff} &  $\sigma${\logg} & $\sigma$[Fe/H] & $\sigma${\vt} \\
    & 100~K & 0.3~dex & 0.3~dex & 0.3~{\kms} & & & 100~K & 0.3~dex & 0.3~dex & 0.3~{\kms} 
 }
\startdata
\ion{Fe}{1} & 0.08 & $-0.01$ & 0.01 & $-0.02$ & $-0.08$ & & 0.07 & $-0.01$ & 0.01 & $-0.03$ \\
\ion{Fe}{2} & 0.01 & $ 0.10$ & 0.00 & $-0.01$ & $-0.06$ & & 0.02 & 0.10 & 0.00 & $-0.02$ \\
C (CH)      & 0.15 & $-0.10$ & 0.00 & 0.00    & $-0.10$ & & 0.15 & $-0.12$ & 0.01 & 0.00 \\
\ion{Na}{1} & 0.11 & $-0.15$ & 0.00 & $-0.05$ & $-0.11$ & & 0.07 & $-0.03$ & 0.00 & $-0.07$ \\
\ion{Mg}{1} & 0.04 & $-0.08$ & $-0.02$ & $-0.05$ &$-0.13$ & & 0.06 & $-0.04$ & 0.00 & $-0.06$ \\
\ion{Ca}{1} & 0.05 & $0.00$ & $0.00$ & $-0.01$ & $-0.08$ & & 0.05 & $-0.01$ & 0.00 & $-0.01$ \\
\ion{Ti}{1} & 0.08 & $0.00$ & $0.01$ & $0.00$  & $-0.08$ & & 0.07 & $-0.01$ & 0.01 & $0.00$ \\
\ion{Ti}{2} & 0.04 & $0.10$ & $0.00$ & $-0.02$ & $-0.06$ & & 0.04 & $0.09$ & 0.00 & $-0.02$ \\
\ion{Cr}{1} & 0.08 & $0.00$ & $0.01$ & $-0.01$ & $-0.08$ & & 0.08 & $-0.01$ & 0.01 & $-0.02$ \\
\ion{Ni}{1} & 0.08 & $0.01$ & $0.01$ & $-0.01$ & $-0.07$ & & 0.07 & $-0.01$ & 0.01 & $-0.01$ \\
\ion{Sr}{2} & 0.08 & $0.03$ & $0.00$ & $-0.16$ & $-0.10$ & & 0.07 & $0.02$ & $-0.01$ & $-0.2$ \\
\ion{Ba}{2} & 0.09 & $-0.02$ & $0.00$ & $-0.10$ & $-0.14$ & & 0.07 & $0.02$ & $-0.02$ & $-0.15$ \\
\ion{La}{2} & 0.06 & $0.10$ & $0.00$ & $-0.02$ & $-0.06$ & & \nodata & \nodata & \nodata & \nodata \\
\ion{Eu}{2} & 0.06 & $0.11$ & $0.00$ & $0.02$ & $-0.05$ & & \nodata & \nodata & \nodata & \nodata \\
\ion{Pb}{1} & 0.08 & $0.01$ & $0.00$ & $-0.01$ & $-0.10$ & & 0.07 & $-0.01$ & $0.01$ & $-0.02$ 
\enddata
\end{deluxetable}

\begin{deluxetable}{lccccccccccc}
 \tablewidth{0pt}
\tablecaption{COMPARISONS OF ATMOSPHERIC PARAMETERS AND ABUNDANCES WITH SDSS ESTIMATES \label{tab:comp}}
\tablehead{
    & \multicolumn{2}{c}{{\teff}} && \multicolumn{2}{c}{{\logg}} && \multicolumn{2}{c}{[Fe/H]} && \multicolumn{2}{c}{[C/Fe]}\\
     \cline{2-3}\cline{5-6}\cline{8-9}\cline{11-12}
    & SDSS & this work            && SDSS & this work            && SDSS & this work           && SDSS & this work 
 }
\startdata
{\obja} & 6595 & 6500 && 3.58 & 4.5 &&  $-2.49$  & $-2.41$ &&   2.50 & 2.3 \\
{\objc} & 6970 & 6600 && 4.01 & 4.1 &&  $-2.68$  & $-3.11$ &&   2.71 & 2.9 \\
{\objd} & 6213 & 6300 && 3.28 & 4.0: &&  $-2.88$  & $-3.16$ &&   1.19 & $<2.2$  \\
{\obje} & 6264 & 6200 && 3.67 & 4.0 &&  $-2.65$  & $-2.51$ &&   2.58 & 2.7 \\
{\objf} & 6656 & 6700 && 3.19 & 4.2 &&  $-2.44$  & $-2.52$ &&   2.27 & 2.1: \\
{\objb} & 6489 & 6600 && 3.76 & 4.5 &&  $-2.22$  & $-2.05$ &&   1.93 & 2.0 
\enddata
\end{deluxetable}

\clearpage
\begin{deluxetable}{lccccccccccccccccc}
\tabletypesize{\scriptsize}
\rotate
\tablewidth{0pt}
\tablecaption{\label{tab:cemp_to} ABUNDANCES RESULTS}
\tablehead{
            & [Fe/H]& [C/Fe] & [N/Fe] & [Na/Fe] &[Mg/Fe] &[Ca/Fe] &[Sc/Fe] &[Ti/Fe] &[Cr/Fe] &[Ni/Fe] &[Zn/Fe]& [Sr/Fe] &[Ba/Fe]& [Pb/Fe] & Teff & logg & ref.\tablenotemark{a}
}
\startdata
{\LP}       & $-$2.53 & 2.14 & 1.20 & 0.60 & 0.42 & 0.22 & 0.20 & 0.47 &$-$0.07 & 0.07  &0.00 & 0.08 & 2.08 & 2.53 &  6300 & 4.30  & 1  \\ 
{\CS}       & $-$2.06 & 2.07 & 1.40 & 0.14 & 0.22 & 0.19 &\nodata & 0.39 &$-$0.05 & 0.35 &\nodata & 0.77 & 2.39 & 3.16 &  6800 & 4.10 & 1     \\ 
{\obja}     & $-$2.41 & 2.32 &\nodata & 0.78 & 0.25 & 0.26 &\nodata & 0.51 &$-$0.18 &$-$0.17 &\nodata &$-$0.27 & 0.29 &\nodata &  6500 & 4.50 & 1    \\ 
{\objc}     & $-$3.11 & 2.92 &\nodata & 0.69 & 0.61 &\nodata &\nodata & 0.66 & 0.62 &\nodata &\nodata & 1.35 & 2.75 & 3.41 &  6600 & 4.10 & 1 \\ 
{\obje}     & $-$2.51 & 2.72 &\nodata & 1.31 & 0.52 & 0.28 & 0.35 & 0.40 &$-$0.29 & 0.05 &\nodata & 0.60 & 1.81 & 3.10 &  6200 & 4.00 & 1 \\ 
{\objf}     & $-$2.52 & 2.1  &\nodata & 2.71 & 1.13 & 0.79 &\nodata & 0.60 &\nodata &\nodata &\nodata & 2.25 & 3.40 &\nodata &  6700 & 4.20  & 1 \\       
{\objb}     & $-$2.05 & 2.   &\nodata & 0.33 & 0.27 & 0.32 &\nodata & 0.62 &$-$0.10 & 0.08 &\nodata & 0.68 & 1.50 &\nodata &  6600  & 4.50 & 1 \\ 
CS29528-028 & $-$2.86 & 2.77 &\nodata & 2.33 & 1.69 & 0.46 & 0.59 & 0.87 &\nodata & 0.26 &\nodata &\nodata & 3.27 &\nodata &  6800 & 4.00  & 2 \\ 
CS22898-027 & $-$2.26 & 2.20 & 0.90 & 0.33 & 0.41 & 0.40 &\nodata & 0.41 &$-$0.10 & 0.02 & 0.92 & 0.92 & 2.23 & 2.84 &  6250 & 3.70  & 3,4 \\  
CS29497-030 & $-$2.57 & 2.47 & 2.12 & 0.58 & 0.44 & 0.47 & 0.67 & 0.64 & 0.03 & 0.04 &\nodata & 0.84 & 2.32 & 3.55 &  7000 & 4.10  & 5 \\  
HE2148-1247 & $-$2.32 & 1.91 & 1.65 &\nodata & 0.50 & 0.45 & 0.59 & 0.55 &$-$0.35 & 0.06 &\nodata & 0.76 & 2.36 & 3.12 &  6380 & 3.90   & 6 \\ 
HE0024-2523 & $-$2.72 & 2.60 & 2.10 &$-$0.17 & 0.73 & 0.66 & 0.37 & 0.85 &$-$0.41 &\nodata &\nodata & 0.34 & 1.46 & 3.30 &  6625 & 4.30    & 7 \\ 
HE22881-036 & $-$2.06 & 1.96 & 1.00 & 0.16 & 0.40 & 0.62 &\nodata & 0.33 &\nodata &\nodata &\nodata & 0.59 & 1.93 &\nodata &  6200 & 4.00   & 8 \\  
HE0007-1832 & $-$2.65 & 2.55 & 1.85 &\nodata & 0.76 & 0.32 &\nodata & 0.39 &\nodata & 0.02 &\nodata &\nodata & 0.16 &\nodata &  6515 & 3.80    & 9\\  
HE0338-3945 & $-$2.42 & 2.13 & 1.55 & 0.36 & 0.30 & 0.38 & 0.53 & 0.37 &$-$0.12 & 0.01 &\nodata & 0.74 & 2.41 & 3.10 &  6160 & 4.13   & 10 \\   
HE1105+0027 & $-$2.42 & 2.00 &\nodata &\nodata & 0.47 & 0.47 & 0.28 & 0.32 & 0.05 &$-$0.29 &\nodata &\nodata & 2.45 &\nodata &  6132 & 3.50  & 11  \\   
HE0143-0441 & $-$2.31 & 1.98 & 1.73 &\nodata & 0.63 & 0.43 & 0.67 & 0.40 &$-$0.38 &$-$0.31 & 0.46 & 0.86 & 2.32 & 3.11 &  6240 & 3.70   & 12 \\   
CS31080-095 & $-$2.80 & 2.69 & 0.70 &$-$0.48 & 0.65 & 0.17 &$-$0.02 & 0.32 & 0.02 & 0.09 & 0.58 &$-$0.41 & 0.77 &\nodata &  6050 & 4.50  & 13\\ 
CS22958-042 & $-$2.80 & 3.15 & 2.15 & 2.62 & 0.32 & 0.36 & 0.05 & 0.32 &$-$0.15 &$-$0.09 &\nodata &$-$0.20 &\nodata &\nodata &  6250 & 3.50  & 13\\ 
CS29528-041 & $-$3.25 & 1.59 & 3.00 & 1.00 & 0.40 & 0.40 & 0.26 & 0.40 &$-$0.17 & 0.00 &\nodata &$-$0.20 & 0.97 &\nodata &  6150 & 4.00  & 13\\ 
\enddata
\tablenotetext{a}{References-- (1)This work; (2)\citet{aoki07};
(3)\citet{aoki02b}; (4)\citet{aoki02c}; (5)\citet{ivans05};
(6)\citet{cohen03}; (7)\citet{lucatello03}; (8)\citet{preston01};
(9)\citet{cohen04}; (10)\citet{jonsell06}; (11)\citet{barklem05};
(12)\citet{cohen06}; (13)\citet{sivarani06}}
\end{deluxetable}

\begin{deluxetable}{lrrrrrrrrrrrrrrrrr} 
 \tablewidth{0pt}
\rotate
\tabletypesize{\footnotesize}
\tablecaption{KINEMATICS DATA \label{tab:kinematics}}
\tablehead{
star    & $D$ & $V_{\rm helio}$ & $\mu_{\alpha}$ & $\mu_{\delta}$ & & $U$ & $\sigma(U)$ & $V$ & $\sigma(V)$ & $W$ & $\sigma(W)$ & $V_{\phi}$ & $\sigma(V_{\phi})$ & $e$ & $r_{\rm min}$ & $r_{\rm max}$ & $Z_{\rm max}$ \\
\cline{4-5} \cline{7-14}
        & (kpc) & ({\kms}) & \multicolumn{2}{c}{(mas/yr)} & & \multicolumn{8}{c}{({\kms})} &  & (kpc) & (kpc) & (kpc)
}
\startdata
{\obja} &  2.27 & $-146$ & $-11$ & $-50$ && $-399$ & 48 & $-403$ & 47 & $   2$ & 17 & $-154$ & 43 & 0.92 & 3 & 66 &  22 \\
{\objc} &  1.75 & $-272$ &  $15$ & $ 11$ && $  15$ & 23 & $ -99$ & 19 & $ 298$ & 13 & $ 120$ & 19 & 0.41 & 9 & 22 &  20 \\
{\objd} &  1.46 & $   2$ &  $21$ & $-10$ && $ -95$ & 13 & $ -78$ & 20 & $ 111$ & 19 & $ 139$ & 20 & 0.38 & 5 & 12 &   5 \\
{\obje} &  1.39 & $-366$ & $-31$ & $-46$ && $-130$ & 18 & $-293$ & 35 & $-392$ & 18 & $ -73$ & 35 & 0.74 & 8 & 54 &  53 \\
{\objf} &  2.43 & $  37$ &  $-2$ & $  1$ && $   5$ & 33 & $  30$ & 20 & $  46$ & 27 & $ 242$ & 20 & 0.24 & 8 & 13 &   2 \\
{\objb} &  2.23 & $-418$ &  $20$ & $-33$ && $ 209$ & 22 & $-510$ & 32 & $-138$ & 41 & $-326$ & 32 & 0.66 & 6 & 31 &   9 
\enddata
\end{deluxetable}

\clearpage
\pagestyle{plaintop}

\begin{thebibliography}{}




\bibitem[Adelman-McCarthy et al.(2007)]{adelman-mccarthy07} 
Adelman-McCarthy, J.~K., et al.\ 2007, \apjs, 172, 634 

\bibitem[Alonso, Arribas, \& Mart\'{i}nez-Roger (1996)]{alonso96}
  Alonso, A., Arribas, S., \& {Mart{\'\i}nez-Roger}, C. 1996, \aap,
  313, 873

\bibitem[Alvarez \& Plez (1998)]{alvarez98} Alvarez, R., \& Plez,
B.\ 1998, \aap, 330, 1109

\bibitem[Aoki et al.(2005)]{aoki05} Aoki, W., et al.\ 2005, 
\apj, 632, 611 

\bibitem[Aoki et al.(2007)]{aoki07} Aoki, W., Beers, T.~C., 
Christlieb, N., Norris, J.~E., Ryan, S.~G., \& Tsangarides, S.\ 2007, \apj, 
655, 492 

\bibitem[Aoki et al.(2002a)]{aoki02a} Aoki, W., Norris, J.~E., 
Ryan, S.~G., Beers, T.~C., \& Ando, H.\ 2002a, \apj, 567, 1166 

\bibitem[Aoki et al.(2002b)]{aoki02b} Aoki, W., Norris, J.~E., 
Ryan, S.~G., Beers, T.~C., \& Ando, H.\ 2002b, \pasj, 54, 933 

\bibitem[Aoki et al.(2002c)]{aoki02c} Aoki, W., Ryan, S.~G., 
Norris, J.~E., Beers, T.~C., Ando, H., \& Tsangarides, S.\ 2002c, \apj, 580, 
1149 



\bibitem[Asplund(2005a)]{asplund05a} Asplund, M.\ 2005a, \araa, 43, 
481 

\bibitem[Asplund et al.(2005b)]{asplund05b} Asplund, M., Grevesse, 
N., \& Sauval, A.~J.\ 2005b, ASP Conf.~Ser.~336: Cosmic Abundances as 
Records of Stellar Evolution and Nucleosynthesis, 336, 25 

\bibitem[Barklem \& Aspelund-Johansson(2005)]{barklem05} Barklem,
P.~S., \& Aspelund-Johansson, J.\ 2005, \aap, 435, 373

\bibitem[Barklem \& O'Mara(1998)]{barklem98} Barklem, P.~S., \&
O'Mara, B.~J.\ 1998, \mnras, 300, 863

\bibitem[Beers \& Christlieb (2005)]{beers05} Beers, T. C., \&
Christlieb, N. 2005, ARA\&A, 43, 531

\bibitem[Beers et al.(1992)]{beers92} Beers, T.~C., Preston, 
G.~W., \& Shectman, S.~A.\ 1992, \aj, 103, 1987 

\bibitem[Beers et al. (2000)]{beers00} Beers, T. C., et al. 2000, \aj, 119, 2866

\bibitem[Beers et al. (2006)]{beers06} Beers, T.C., et al. 2006, BAAS
38, 168.08

\bibitem[Beers et al.(2007)]{beers07} Beers, T.~C., et al.\ 
2007, \apjs, 168, 128 

\bibitem[Bond (1974)]{bond74} Bond, H.~E.\ 1974, \apj, 194, 95

\bibitem[Busso et al.(2001)]{busso01} Busso, M., Gallino, R., 
Lambert, D.~L., Travaglio, C., \& Smith, V.~V.\ 2001, \apj, 557, 802 

\bibitem[Busso et al.(1999)]{busso99} Busso, M., Gallino, R., 
\& Wasserburg, G.~J.\ 1999, \araa, 37, 239

\bibitem[Carollo et al.(2007)]{carollo07} Carollo, D., et al.\ 
2007, \nat, 450, 1020 

\bibitem[Castelli \& Kurucz(2003)]{castelli03} Castelli, F., \& 
Kurucz, R.~L.\ 2003, Modelling of Stellar Atmospheres, 210, 20P 


\bibitem[Charbonnel et al. (2007)]{charbonnel07} Charbonnel, C., \& Zahn, J.-P. 2007, \aap, 467, L15


\bibitem[Christlieb et al.(2002)]{christlieb02} Christlieb, N., et 
al.\ 2002, \nat, 419, 904 

\bibitem[Cohen et al.(2003)]{cohen03}  Cohen, J.~G., Christlieb, 
N., Qian, Y.-Z., \& Wasserburg, G.~J.\ 2003, \apj, 588, 1082 

\bibitem[Cohen et al.(2004)]{cohen04} Cohen, J.~G., et al.\ 
2004, \apj, 612, 1107 

\bibitem[Cohen et al.(2006)]{cohen06} Cohen, J.~G., et al.\ 
2006, \aj, 132, 137 

\bibitem[Collet et al.(2005)]{collet05} Collet, R., Asplund, M., 
\& Th{\'e}venin, F.\ 2005, \aap, 442, 643

\bibitem[Collet et al.(2006)]{collet06} Collet, R., Asplund, M., 
\& Trampedach, R.\ 2006, \apjl, 644, L121 

\bibitem[Cui \& Zhang(2006)]{cui06} Cui, W., \& Zhang, B.\ 
2006, \mnras, 368, 305 

\bibitem[Denissenkov \& Pinsonneault(2007)]{denissenkov07} 
Denissenkov, P.~A., \& Pinsonneault, M.\ 2007, ArXiv e-prints, 709, 
arXiv:0709.4240 

\bibitem[Downes et al. (2004)]{downes04} Downes, R.A., et al. 2004,
\aj, 127, 2838

\bibitem[Frebel et al.(2005)]{frebel05} Frebel, A., et al.\ 
2005, \nat, 434, 871 

\bibitem[Frebel et al. (2006)]{frebel06} Frebel, A., et al. 2006,
\apj, 652, 1585

\bibitem[Fukugita et al. (1996)]{fukugita96} Fukugita, M., Ichikawa, T., Gunn, J.E., Doi, M., Shimasaku, K., \&
Schneider, D.P. 1996, \aj, 111, 1748  

\bibitem[Gunn et al. (1998)]{gunn98} Gunn, J.E., et al. 1998, \aj, 116, 3040 

\bibitem[Gunn et al. (2006)]{gunn06} Gunn, J.E., et al. 2006, \aj, 131, 2332 

\bibitem[Hogg et al. (2001)]{hogg01} Hogg, D.W., Finkbeiner, D.P., Schlegel, D.J., \& Gunn, J.E. 2001,
   \aj, 122, 2129 

\bibitem[Ivans et al.(2005)]{ivans05} Ivans, I.~I., Sneden, C., 
Gallino, R., Cowan, J.~J., \& Preston, G.~W.\ 2005, \apjl, 627, L145 

\bibitem[ Ivez\'ic et al. (2004)]{ivezic04} Ivez\'ic, Z., et al. 2004, AN, 325, 583 

\bibitem[Jonsell et al. (2006)]{jonsell06} Jonsell, K., Barklem, P. S.,
Gustafsson, B., Christlieb, N., Hill, V., Beers, T. C., \& Holmberg, J.
2006, A\&A 451, 651

\bibitem[Keenan (1942)]{keenan42} Keenan, P. C., \apj, 96, 101



\bibitem[Kim et al.(2002)]{y2} Kim, Y.-C., Demarque, P., 
Yi, S.~K., \& Alexander, D.~R.\ 2002, \apjs, 143, 499 

\bibitem[Komiya et al.(2007)]{komiya07} Komiya, Y., Suda, T., 
Minaguchi, H., Shigeyama, T., Aoki, W., \& Fujimoto, M.~Y.\ 2007, \apj, 
658, 367 

\bibitem[Kupka et al.(1999)]{kupka99} Kupka, F., Piskunov, N.,
Ryabchikova, T.~A., Stempels, H.~C., \& Weiss, W.~W.\ 1999, \aaps,
138, 119

\bibitem[{Kurucz(1993)}]{kurucz93} Kurucz, R.~L. 1993, CD-ROM 13,
ATLAS9 Stellar Atmospheres Programs and 2~km/s Grid (Cambridge:
Smithsonian Astrophys. Obs.)

\bibitem[Lawler et al.(2001)]{lawler01} Lawler, J.~E., 
Bonvallet, G., \& Sneden, C.\ 2001, \apj, 556, 452 

\bibitem[Lee et al. (2007a)]{lee07a} Lee, Y.S., et al. 2007a, \aj,
submitted

\bibitem[Lee et al. (2007b)]{lee07b} Lee, Y.S., et al. 2007b, \aj,
submitted

\bibitem[Lucatello et al.(2006)]{lucatello06} Lucatello, S., Beers, 
T.~C., Christlieb, N., Barklem, P.~S., Rossi, S., Marsteller, B., Sivarani, 
T., \& Lee, Y.~S.\ 2006, \apjl, 652, L37 

\bibitem[Lucatello et al.(2003)]{lucatello03} Lucatello, S., Gratton,
R., Cohen, J.~G., Beers, T.~C., Christlieb, N., Carretta, E., \&
Ram{\'{\i}}rez, S.\ 2003, \aj, 125, 875

\bibitem[Lucatello et al.(2005)]{lucatello05} Lucatello, S., 
Tsangarides, S., Beers, T.~C., Carretta, E., Gratton, R.~G., \& Ryan, 
S.~G.\ 2005, \apj, 625, 825 

\bibitem[Lupton et al.(2001)]{lupton01} Lupton, R., Gunn, J.~E., 
Ivezi{\'c}, Z., Knapp, G.~R., \& Kent, S.\ 2001, Astronomical Data Analysis 
Software and Systems X, 238, 269 

\bibitem[Margon et al. (2002)]{margon02} Margon, B., et al. 2002, \aj, 124, 1651 

\bibitem[Marsteller (2007)]{marsteller07} Marsteller, B. 2007, PhD Thesis,
Michigan State University

\bibitem[Marstellar et al. (2006)]{marstellar06} Marsteller, B., et
al. 2006, BAAS, 38, 242.02

\bibitem[Masseron et al.(2006)]{masseron06} Masseron, T., et al.\ 
2006, \aap, 455, 1059 


\bibitem[{McWilliam(1998)}]{mcwilliam98}
McWilliam, A. 1998, \aj, 115, 1640


\bibitem[Munn et al. (2004)]{munn04} Munn, J. A., et al. 2004, \aj, 127, 3034

\bibitem[{Munari \& Zwitter(1997)}]{munari97}
Munari, U., \& Zwitter, T. 1997, A\&A, 318, 269

\bibitem[{Noguchi {et~al.}(2002)Noguchi, Aoki, \& et~al.}]{noguchi02}
Noguchi, K. et~al. 2002, PASJ, 54, 855


\bibitem[Norris et al.(1997a)]{norris97a} Norris, J.~E., Ryan, 
S.~G., \& Beers, T.~C.\ 1997a, \apj, 488, 350 

\bibitem[Norris et al.(1997b)]{norris97b} Norris, J.~E., Ryan, 
S.~G., \& Beers, T.~C.\ 1997b, \apjl, 489, L169 

\bibitem[{Norris {et~al.}(2001)Norris, Ryan, \& Beers}]{norris01}
Norris, J. E., Ryan, S. G., \& Beers, T. C. 2001, ApJ, 561, 1034


\bibitem[Norris et al.(2007)]{norris07} Norris, J.~E., 
Christlieb, N., Korn, A.~J., Eriksson, K., Bessell, M.~S., Beers, T.~C., 
Wisotzki, L., \& Reimers, D.\ 2007, \apj, 670, 774 

\bibitem[Preston \& Sneden(2001)]{preston01} Preston, G.~W., \& 
Sneden, C.\ 2001, \aj, 122, 1545 

\bibitem[Pier et al. (2003)]{pier03} Pier, J.R., Munn, J.A., Hindsley, R.B., Hennessy, G.S., Kent, S.M.,
Lupton, R.H., \& Ivez\'ic, Z. 2003, \aj, 125, 1559 

\bibitem[Plez \& Cohen(2005)]{2005A&A...434.1117P} Plez, B., \& Cohen,
J.~G.\ 2005, \aap, 434, 1117

\bibitem[Reyniers et al.(2004)]{reyniers04} Reyniers, M., Van 
Winckel, H., Gallino, R., \& Straniero, O.\ 2004, \aap, 417, 269 

\bibitem[{Schlegel {et~al.}(1998)Schlegel, Finkbeiner, \&
  Davis}]{schlegel98}
Schlegel, D., Finkbeiner, D., \& Davis, M. 1998, ApJ, 500, 525

\bibitem[Smith et al. (2002)]{smith02} Smith, J.A., et al. 2002, \aj, 123, 2121 

\bibitem[Sneden et al.(2003)]{sneden03} Sneden, C., et al.\ 
2003, \apj, 591, 936 

\bibitem[Simons et al. (1989)]{simons89} Simons, J.W., Palmer, B.A.,
Hof, D.E., \& Oldenborg, R.C. 1989, J. Opt. Soc. Am. B., 6, 1097

\bibitem[Sivarani et al.(2006)]{sivarani06} Sivarani, T., et al.\ 
2006, \aap, 459, 125 

\bibitem[Skrutskie et al.(2006)]{skrutskie06} Skrutskie, M.~F., et 
al.\ 2006, \aj, 131, 1163 

\bibitem[Stancliffe et al.(2007)]{stancliffe07} Stancliffe, R.~J., 
Glebbeek, E., Izzard, R.~G., \& Pols, O.~R.\ 2007, \aap, 464, L57 

\bibitem[Stoughton et al. (2002)]{stoughton02} Stoughton, C., et al. 2002, \aj, 123, 485 


\bibitem[Thompson et al.(2007)]{thompson08} Thompson, I.~B., et 
al.\ 2007, ApJ, in press, ArXiv e-prints, 712, arXiv:0712.3228 

\bibitem[Tucker et al. (2006)]{1234} Tucker, D., et al. 2006, AN, 327, 821 

\bibitem[Tumlinson(2007)]{tumlinson07} Tumlinson, J.\ 2007, \apjl, 
664, L63 

\bibitem[van den Hoek \& Groenewegen(1997)]{vandenhoek97} van den
Hoek, L.~B., \& Groenewegen, M.~A.~T.\ 1997, \aaps, 123, 305

\bibitem[York et al. (2000)]{york00} York, D.G., et al. 2000, \aj, 120, 1579
 
\bibitem[Zhao \& Newberg (2006)]{zhao06} Zhao, C., \& Newberg, H.J. 2006, unpublished manuscript
(astro-ph/0612034)



\end{thebibliography}
\end{document}